\documentclass[11pt,nofootinbib,preprintnumbers,amsmath,amssymb,tightenlines]{revtex4}
\usepackage[colorlinks,
            linkcolor=blue,
            anchorcolor=blue,
            citecolor=blue]{hyperref}
\usepackage{natbib}
\usepackage{doi}
\usepackage{url}
\usepackage{exscale}
\usepackage{amsmath}
\usepackage{booktabs}
\usepackage{graphicx}
\usepackage{exscale}
\usepackage{relsize}
\usepackage{graphicx}
\usepackage{dcolumn}
\usepackage{bm}
\usepackage{multirow}
\usepackage{CJK}
\usepackage{diagbox}
\usepackage{subfigure}
\usepackage{slashed}
\usepackage{makecell}
\usepackage{adjustbox}
\usepackage{longtable}
\usepackage{multirow}
\usepackage{amsmath}
\makeatletter

\newcommand{\Rmnum}[1]{\expandafter\@slowromancap\romannumeral #1@}
\def \nl {\nonumber\\}
\makeatother

\begin{document}

\title{Two-body nonleptonic decays of the heavy mesons in the factorization approach}

\author{Shuo-Ying Yu}
\author{Xian-Wei Kang}
\affiliation{Key Laboratory of Beam Technology of the Ministry of Education,
College of Nuclear Science and Technology, Beijing Normal University, Beijing 100875, China,\\
and Institute of Radiation Technology, Beijing Academy
of Science and Technology, Beijing 100875, China }

\author{V. O. Galkin}
\affiliation{Federal Research Center ``Computer Science and Control'', Russian Academy of Sciences,  Vavilov Street 40, 119333 Moscow, Russia}

\begin{abstract}
In the framework of the factorization approach we calculate the branching fractions of 100 two-body nonleptonic decay channels in total, including 44 channels of the charm meson decays and 56 channels of the bottom meson decays. For charm meson decays, we test and confirm the previous observation that taking the limit for the number of colors $N\to\infty$ significantly improves theoretical predictions. For bottom meson decays, the penguin contributions are included in addition. As an essential input, we employ the weak decay form factors obtained in the framework of the relativistic quark model based on the quasi-potential approach. These form factors have well been tested by calculating observables in the semileptonic $D$ and $B$ meson decays and confronting obtained results with experimental data. In general, the predictions for the nonleptonic decay branching fractions are acceptable. However, for a quantitative calculation it is necessary to account for a more subtle effects of the final-state interaction.
\end{abstract}

\maketitle

\section{Introduction}
Nonleptonic decays of the heavy mesons offer an environment to
understand the nature of quantum chromodynamics (QCD). Experimentally many such decays have been measured. Theoretically nonleptonic decays involve more complex mechanism than the leptonic and semileptonic ones due to the local four-quark operators. A usual treatment is the factorization approach, where the decay amplitude is factorized into the product of the meson decay constant and weak transition form factors. An intuitive
justification for the factorization approximation comes from the so-called colour transparency proposed by Bjorken \cite{Bjorken:1988kk}, where for the energetic $B$ decay, the final light meson flies very fast in the opposite direction to the other meson, thus almost escaping the color field of the parent particle. As a result, the factorization holds. However, the strong and complicated final state interaction does challenge the factorization approximation. And this part is very hard to be quantified, therefore, in this paper we test how factorization works for the nonleptonic decays.

The form factors embodying the dynamics of the meson weak transitions are an essential input.  As a motivation and also a new point of this paper,
we adopt the form factors which are derived from the relativistic quark model based on the quasi-potential approach. The numerical values of form factor parameters can be found in Refs.~\cite{Faustov:2019mqr} and \cite{Faustov:2022ybm}, containing the results for the weak $D$ and $B$ decays to the pseudoscalar and/or vector mesons in the final states. In this relativistic quark model, the meson wave functions are explicitly obtained as numerical solutions of the relativistic Schr\"odinger-like bound-state equation and not assumed to be an empirical Gaussian function. Moreover,
no free parameters are involved since they have been fixed by the previous studies of hadron spectroscopy. All relativistic effects including the transformation of meson wave functions from the rest reference frame to the moving one, and the contribution of intermediate negative energy states are included. It is important to note that the form factors are predicted in the whole kinematically allowed region. The values of these form factors have been well tested confronting with experiment by a series of the calculated semileptonic decay observables, e.g., the branching fractions, forward-backward asymmetries and polarizations. Similar work concerning the application of those form factors to the nonleptonic decays can be found in Refs.~\cite{Ebert:1997mg,Faustov:2012mt,Faustov:2013ima}. A very recent study of the charmless two-body $B$ meson decays is performed in the perturbative QCD factorization approach \cite{Chai:2022kmq} as a more advanced tool. See also Refs.~\cite{Ali:2007ff,Yang:2020xal,Xiao:2011tx} for some earlier works.

In this paper we calculate the branching fractions of the charm and bottom meson nonleptonic decays in the framework of the factorization approach based on the effective weak Hamiltonian. Experimental results from PDG and other theoretical predictions are also compiled for a direct comparison. For charm meson decays, the penguin contributions are highly suppressed, thus we neglect them. We have considered the cases of the color numbers $N=3$ as in reality and $N\to\infty$ \footnote{The application of large $N_c$ approach to the hadron physics can be seen in e.g. Refs.~\cite{Dai:2017uao,Manohar:1998xv,Guo:2012yt}}. In the latter case, the discrepancy between the theory and experimental results are expected to be significantly reduced due to
the experience in 1980s \cite{Tadic:1982vn,Bauer:1984zv,Buras:1985xv,Halperin:1994hg,Cheng:2018hwl}. For the decay process of the $B$ mesons, we consider both tree-level and penguin loop-level contributions. The latter can be as large as the former, or even dominant.

This paper is organized as follows.  In Sec.~\ref{sec:II} and Sec.~\ref{sec:III} we briefly describe the effective Hamiltonian governing the $D$ and $B$ weak decays. In Sec.~\ref{sec:IV} we collect the input values. In Sec.~\ref{sec:V} we show our numerical results and discuss them. Some care should also be taken for the conventions for the definitions of the decay constants and form factors. Conclusion are given in Sec.~\ref{sec:VI}.

\section{Factorization in the charm meson two-body decays}
\label{sec:II}
In the standard model, the effective Hamiltonian of the weak charm meson decay process reads
\begin{eqnarray}
H_{\text{eff}}=\frac{G_F}{\sqrt{2}}V_{cq_1}^*V_{u{q_2}}[c_1(\mu)(\bar{q}_{1\alpha} c_\alpha)_{V-A}(\bar{u}_\beta q_{2\beta})_{V-A}+c_2(\mu)(\bar{q}_{1\alpha} c_\beta)_{V-A}(\bar{u}_\beta q_{2\alpha})_{V-A}]
\end{eqnarray}
where both $q_{1, 2}$ can be either $s$ or $d$ quarks, $(\bar{q}_1q_2)_{V-A}=\bar{q}_1\gamma_\mu(1-\gamma_5)q_2$ and $G_F=1.166\times10^{-5} \text{GeV}^{-2}$ is the Fermi coupling constant; $\alpha$ and $\beta$ are color indexes; $c_1(\mu)$ and $c_2(\mu)$ are Wilson coefficients for which we use the values $c_1=1.26,c_2=-0.51$ \cite{Kamal:1995fr} at the scale $\mu=m_c$. The penguin contributions are tiny and thus can be ignored.

Using the Fierz identity ($i, j, k, l$ are color indices)
\begin{equation}
\delta_{lj}\delta_{ik}=\frac{1}{N}\delta_{ij}\delta_{lk}+2T_{lk}^aT_{ij}^a,
\end{equation}
with $T^a=\frac{\lambda^a}{2}$, and $\lambda^a$ ($a=1, 2,\cdots, 8$) being the Gell-Mann matrices, and $N$ being the number of colors, one has \begin{equation}\label{eq:arrange}
 (\overline{s}s)_{V-A}(\overline{u}c)_{V-A}=\frac{1}{N}(\overline{s}c)_{V-A}(\overline{u}s)_{V-A}+2(\overline{s}_\alpha T_{\beta\rho}s_\rho)_{V-A}(\overline{u}_\beta T_{\alpha\sigma}c_\sigma)_{V-A}.
\end{equation}
In the factorization approximation the second term of Eq.~\eqref{eq:arrange}, which contributes to the nonfactorizable part, is neglected. Then we can write the effective Hamiltonian corresponding to the color-favored decay process
\begin{equation}
 H_\text{cf}=\frac{G_F}{\sqrt{2}}V_{cq_1}^*V_{uq_2}a_1(\bar{q}_1c)_{V-A}(\bar{u}q_2)_{V-A},
\end{equation}
and for the color-suppressed case
\begin{equation}
H_\text{cs}=\frac{G_F}{\sqrt{2}}V_{cq_1}^*V_{uq_2}a_2(\bar{q}_1q_2)_{V-A}(\bar{u}c)_{V-A},
\end{equation}
with $a_1=c_1+\frac{c_2}{N}, a_2=c_2+\frac{c_1}{N}$. Empirically, taking the choice of $N\rightarrow \infty$ will generally improve theoretical predictions for the charm meson decays, as it was already mentioned in Introduction.  In this sense, part of the nonfactorizable effects have been compensated by the choice of $N\to\infty$.

In the factorization approach, the hadronic matrix element can be expressed by the product of decay constant and the invariant form factors. The decay constant is defined as the matrix element of the weak current between the vacuum and a pseudoscalar (P) or a vector (V) meson:
\begin{equation}
\begin{split}
&\langle P(p_\mu)|(\overline{q}_1q_2)_{V-A}|0\rangle=-if_{P}p_\mu,~~ \langle V|(\overline{q}_1q_2)_{V-A}|0\rangle=if_{V}m_V \epsilon_\mu^*,\\
\end{split}
\end{equation}
where $f_P$ and $f_V$ are the decay constants of the pseudoscalar and vector meson, respectively; $m_V$ and $\epsilon_\mu$ are the mass and polarization vector of the vector meson.

For the $D\rightarrow P$ transition (with the momenta $p_D,\,p_P$ and masses $m_D,\,m_P$ of the initial and final mesons, respectively), the matrix element of the weak current is parameterized as
\begin{equation}
\begin{split}
&\langle P|(\overline{q}\gamma^\mu c)|D\rangle=f_{+}(q^2)\left[p_{D}^\mu+p_P^\mu-\frac{m_{D}^2-m_P^2}{q^2}q^\mu\right]+ f_{0}(q^2)\frac{m_{D}^2-m_P^2}{q^2}q^\mu,\\
&\langle P|(\overline{q}\gamma^\mu \gamma^5c)|D\rangle=0.\\
\end{split}
\end{equation}
For the $D\rightarrow V$ transition
\begin{equation}
\begin{split}
\langle V|(\overline{q}\gamma^\mu c)|D\rangle&=\frac{2iV(q^2)}{m_{D}+m_V}\varepsilon^{\mu\nu\rho\sigma}\epsilon_\nu^*p_{D\rho}p_{V\sigma},\\
\langle V|(\overline{q}\gamma^\mu\gamma_5 c)|D\rangle&=2m_V A_{0}(q^2)\frac{\epsilon^*\cdot q}{q^2}q^\mu
+(m_{D}+m_V) A_{1}(q^2)\left(\epsilon^{*\mu}-\frac{\epsilon^*\cdot q}{q^2}q^\mu\right)\\&-A_{2}(q^2)\frac{\epsilon^*\cdot q}{m_{D}+m_V}\left[p_{D}^\mu+p_V^\mu-\frac{m_{D}^2-m_V^2}{q^2}q^\mu\right].
\end{split}
\end{equation}
In these equations, $q=p_D-p_{P,V}$ is the four-momentum transfer between the initial $D$ and final $P$ or $V$ mesons.
For the case of the nonleptonic two-body decay considered below, $q$ is just the on-shell momentum of the meson created from vacuum.

The matrix element $X_{D\to M_1,M_2}=\langle M_1|(\overline{q}c)_{V-A}|D\rangle\langle M_2|(\overline{u}q)_{V-A}|0\rangle$ is then simplified as
\begin{equation}
\begin{split}
  &M_2=P: \quad X_{D\rightarrow M_1,P}=-if_Pm_P\epsilon^{\dagger\mu}_{\lambda}\langle M_1|(\overline{q}c)_{V-A}|D\rangle,\\
  &M_2=V: \quad X_{D\rightarrow M_1,V}=if_Vm_V\epsilon^{\dagger\mu}_{\lambda}\langle M_1|(\overline{q}c)_{V-A}|D\rangle,
\end{split}
\end{equation}
where we adopt the convention where the final meson $M_2$ ($P$ or $V$) after the comma in the subscript of $X$ is generated from vacuum.

In the rest frame of the initial $D$ meson, one has the explicit representations of the momentum and polarization vectors:
\begin{equation}
\begin{split}
&p_D^\mu=(m_D,0,0,0),~~p_{M_1}^\mu=(E_1,0,0,|{\bf p}|), ~~q^\mu=(E_2,0,0,-|{\bf p}|),\\
&\epsilon^{\dagger\mu}_t=\frac{1}{\sqrt{q^2}}(E_2,0,0,-|{\bf p}|),~~\epsilon^{\dagger\mu}_\pm=\frac{1}{\sqrt{2}}(0,\pm1,i,0),~~\epsilon^{\dagger\mu}_0=\frac{1}{\sqrt{q^2}}(|{\bf p}|,0,0,-E_2),
\end{split}
\end{equation}
where $E_1$ is the energy of $M_1$ and $E_1+E_2=m_D$, $|{\bf p}|=\lambda^{1/2}(m_D^2, m_1^2, m_2^2)/(2m_D)$ is the momentum of the daughter meson with  $\lambda(x,y,z)=x^2+y^2+z^2-2(xy+yz+xz)$. For convenience, we define the helicity amplitudes
\begin{equation}
H_\lambda\equiv\epsilon^{\dagger\mu}_{\lambda}\langle M_1|(\overline{q}c)_{V-A}|D\rangle,
\end{equation}
with $\lambda=t$ for $M_2=P$ and $\lambda=\pm,0$ for $M_2=V$.

For the process $D\rightarrow P_1,P_2$, one has
\begin{equation} \label{eq:PP}
\begin{split}
&X_{D\to P_1,P_2}=-if_{P_2}m_{P_2}\epsilon^{\dagger\mu}_{t}\langle P_1|(\overline{q}c)_{V-A}|D\rangle\\
&=-if_{P_2}(m_{D}^2-m_{P_1}^2)f_0(m_{P_2}^2)\\
&=-if_{P_2}m_{P_2}H_t;\\
\end{split}
\end{equation}
for $D\rightarrow P,V$,
\begin{equation} \label{eq:PV}
\begin{split}
&X_{D\to P,V}=if_Vm_V\epsilon^{\dagger\mu}_\lambda\langle P|(\overline{q}c)_{V-A}|D\rangle\\
&=i2f_Vf_+(m_V^2)m_{D}|{\bf p}|\\
&=if_Vm_VH_0.\\
\end{split}
\end{equation}
The expressions for $H_0$ and $H_t$ coincide with the ones given in Ref.~\cite{Zhang:2020dla} for the $D\to P$ transition.

For $D\rightarrow V,P$, one has
\begin{equation}\label{eq:VP}
\begin{split}
&X_{D\to V,P}=-if_Pm_P\epsilon^{\dagger\mu}_{t}\langle V|(\overline{q}c)_{V-A}|D\rangle\\
&=2if_PA_0(m_P^2)m_{D}|{\bf p}|\\
&=-if_Pm_PH_t;\\
\end{split}
\end{equation}
for $D\rightarrow V_1,V_2$,
\begin{equation}\label{eq:VV}
\begin{split}
&|X_{D\to V_1,V_2}|^2=f^2_{V_2}m^2_{V_2}(|H_+|^2 + |H_-|^2 + |H_0|^2),\\
&H_\pm=-(m_D+m_{V_1})A_1(m_{V_2}^2)\pm \frac{2m_D|{\bf p}|}{m_D+m_{V_1}}V(m_{V_2}^2),\\
&H_0=-(m_D+m_{V_1})A_1(m_{V_2}^2)\frac{m_D^2-m_{V_1}^2-m_{V_2}^2}{2m_{V_1}m_{V_2}}+\frac{2m_D^2|{\bf p}|^2}{(m_D+m_{V_1})m_{V_1}m_{V_2}}A_2(m_{V_2}^2),\\
\end{split}
\end{equation}
where the definitions of $H_{t}, H_{\pm}, H_0$ coincide with the ones given in Ref.~\cite{Zhang:2020dla} for the $D\to V$ transition.

\section{Factorization of the bottom meson two-body decay amplitudes}
\label{sec:III}
We classify the bottom meson decay channels into two classes according to the effective Hamiltonian. For $\Delta B=\pm1,\Delta C=\pm1$ transitions, e.g., the processes $\bar b\rightarrow \bar{c}u\bar d$, $\bar b\rightarrow\bar{c} u\bar s$, $\bar b\rightarrow \bar{u} c\bar d$, and $\bar b\rightarrow \bar{u} c\bar s$, the effective Hamiltonian reads
 \begin{equation}
\begin{split}
 H_\text{eff}=&\frac{G_F}{\sqrt{2}}[V_{qb}^*V_{q_1q_2}(c_1(\mu)O_1+c_2(\mu)O_2)].
\end{split}
\end{equation}
Then such category is similar to charm decays described above.

For $\Delta B=\pm1,\Delta C=0$ transitions, e.g., $\bar b\rightarrow \bar {c}c\bar d$, $\bar b\rightarrow \bar{c}c\bar s$, $\bar b\rightarrow \bar {u}u \bar d$ and $\bar b\rightarrow \bar {u}u\bar s$, the effective Hamilton reads
\begin{equation}
\begin{split}
 H_\text{eff}=&\frac{G_F}{\sqrt{2}}[V_{q'b}^*V_{q'q}(c_1(\mu)O_1+C_2(\mu)O_2)-V_{tb}^*V_{tq}\sum_{i=3}^{10}c_i(\mu)O_i].
\end{split}
\end{equation}
where $q (q')=s, d, c$;
\begin{equation}
\begin{split}
&O_1=(\overline{b}q')_{V-A}(\overline{q}'q)_{V-A}\,~~~~~~~~~~~~~~~~~~O_2=(\overline{b}_\alpha q'_\beta)_{V-A}(\overline{q}'_\beta q_\alpha)_{V-A},\\
&O_3=(\overline{b}q)_{V-A}\sum_{q'}(\overline{q}'q')_{V-A},~~~~~~~~~~~~~~O_4=(\overline{b}_\alpha q_\beta)_{V-A}\sum_{q'}(\overline{q}'_\beta q'_\alpha)_{V-A},\\
&O_5=(\overline{b}q)_{V-A}\sum_{q'}(\overline{q}'q')_{V+A},~~~~~~~~~~~~~~O_6=(\overline{b}_\alpha q_\beta)_{V-A}\sum_{q'}(\overline{q}'_\beta q'_\alpha)_{V+A},\\
&O_7=\frac{3}{2}(\overline{b}q)_{V-A}\sum_{q'}e_{q'}(\overline{q}'q')_{V+A},~~~~~~~~O_8=\frac{3}{2}(\overline{b}_\alpha q_\beta)_{V-A}\sum_{q'}e_{q'}(\overline{q}'_\beta q'_\alpha)_{V+A},\\
&O_9=\frac{3}{2}(\overline{b}q)_{V-A}\sum_{q'}e_{q'}(\overline{q}'q')_{V-A},~~~~~~~~O_{10}=\frac{3}{2}(\overline{b}_\alpha q_\beta)_{V-A}\sum_{q'}e_{q'}(\overline{q}'_\beta q'_\alpha)_{V-A},\\
\end{split}
\end{equation}
and $e_{q'}$ is the charge of the $q'$ quark.

The even operators $O_{2-10}$ can be rearranged to a color singlet
form by the Fierz transformation
\begin{equation} \label{eq:Fierz}
(\overline{\psi}_1O^i\psi_2)(\overline{\psi}_3O_i\psi_4)=\sum_jC_{ij}(\overline{\psi}_1O^j\psi_4)(\overline{\psi}_3O_j\psi_2),
\end{equation}
where $C_{ij}$ are the Fierz coefficients that are presented in Table \ref{tab:Fierz}. In this way, we have
\begin{equation}
\begin{split}
&O_2=(\overline{b}q)_{V-A}(\overline{q}'q')_{V-A},\\
&O_4=\sum_{q'}(\overline{b}q')_{V-A}(\overline{q}'q)_{V-A},\\
&O_6=-2\sum_{q'}(\overline{b}q')_{S+P}(\overline{q}'q)_{S-P},\\
&O_8=-2\sum_{q'}\frac{3}{2}(\overline{b}q')_{S+P}(\overline{q}'q)_{S-P},\\
&O_{10}=\sum_{q'}\frac{3}{2}(\overline{b}q')_{V-A}e_{q'}(\overline{q}'q)_{V-A}.
\end{split}
\end{equation}
In the above equations, $(\bar q_1 q_2)_{V+A}\equiv \bar q_1 \gamma_\mu(1+\gamma_5)q_2$ and $(\bar q_1 q_2)_{S\pm P}\equiv \bar q_1(1\pm\gamma_5)q_2$.

\begin{table}
\caption{The Fierz coefficients appearing in Eq.~\eqref{eq:Fierz}.}
\label{tab:Fierz}
\begin{ruledtabular}
  \begin{tabular}{lccccc}
$i\backslash j$ &$S$  &$V$  &$T$ &$A$ &$P$\\
  \hline
$V$  &$-1$   &$\frac{1}{2}$  &$0$   &$-\frac{1}{2}$   &$1$\\
$A$  &$-1$   &$-\frac{1}{2}$ &$0$   &$\frac{1}{2}$    &$1$\\
\end{tabular}
\end{ruledtabular}
\end{table}

Here we take $B\rightarrow P_1(q_s q')P_2(q'q)$ ($q_s$ is the spectator quark), as an example, to demonstrate the calculation of the
penguin contribution, with $P_2$ representing a charged pseudoscalar meson. The contribution of $O_4$ therein is proportional to $a_4$ and the contribution of $O_{10}$ is proportional to $\frac{3}{2}e_q'a_{10}$.
The coefficient $a_i$ is related to the Wilson one:
\begin{equation}\label{eq:aici}
\begin{split}
&\text{for}\,\,a_\text{odd},\, a_i=c_i+\frac{c_{i+1}}{N},\\
&\text{for}\,\,a_\text{even},\, a_i=c_i+\frac{c_{i-1}}{N}.\\
\end{split}
\end{equation}
The operator $O_6$ can be further written as
\begin{equation}\label{eq:o6}
\begin{split}
O_6&=-2\sum_{q'}(\overline{b}q')_{S+P}(\overline{q}'q)_{S-P}\\
&=-2[(\overline{b}q')(\overline{q}'q)+(\overline{b}\gamma^5q')(\overline{q}'q)-(\overline{b}q')(\overline{q}'\gamma^5q)
-(\overline{b}\gamma^5q')(\overline{q}'\gamma^5q)].
\end{split}
\end{equation}
Parity conservation leads to
\begin{eqnarray}
\langle P|(\overline{q}_1\gamma^\mu q_2)|0\rangle=0,\quad \langle P|(\overline{q}\gamma^\mu \gamma^5b)|B\rangle=0,
\end{eqnarray}
while equations of motion read (with $m_{1,2}$ being the masses of quarks $q_{1,2}$)
\begin{equation}\label{eq:motion}
  \begin{split}
  &(\overline{q}_1\gamma^5q_2)=\frac{-i}{m_1+m_2}\partial_\mu(\overline{q}_1\gamma^\mu\gamma^5q_2),\quad (\overline{q}_1q_2)=\frac{-i}{m_1-m_2}\partial_\mu(\overline{q}_1\gamma^\mu q_2),
  \end{split}
\end{equation}
and thus only the third term of Eq.~\eqref{eq:o6} survives. Then according to Eq.~(\ref{eq:motion})
\begin{equation}
\begin{split}
\langle P_1|(\overline{b}q')|B\rangle&=\frac{-i}{m_b-m_{q'}}(-iq_\mu)\langle P_1|(\overline{b}\gamma^\mu q')|B\rangle,\\
\langle P_2|(\overline{q}'\gamma^5q)|0\rangle&=\frac{-i}{m_q+m_{q'}}(iq^\mu)\langle P_2|(\overline{b}\gamma^\mu\gamma^5 q')|0\rangle
\end{split}
\end{equation}
and the product is given by
\begin{equation}
\begin{split}
&\langle P_1|(\overline{b}q')|B\rangle\langle P_2|(\overline{q}'\gamma^5q)|0\rangle\\
&=-\frac{m_{P_2}^2}{(m_q+m_{q'})(m_b-m_{q'})}\langle P_1|(\overline{b}\gamma^\mu q')|B\rangle\langle P_2|(\overline{b}\gamma^\mu\gamma^5 q')|0\rangle\\
&=\frac{m_{P_2}^2}{(m_q+m_{q'})(m_b-m_{q'})}X_{B\rightarrow P_1P_2}.
\end{split}
\end{equation}
Therefore, the contribution of $O_6$ is proportional to $\frac{2m_{P_2}^2}{(m_q+m_{q'})(m_b-m_{q'})}a_6$.  And similarly, the contribution of $O_8$ is proportional to $\frac{3}{2}e_q'\frac{2m_{P_2}^2}{(m_q+m_{q'})(m_b-m_{q'})}$. In Table~\ref{tab:penguin} we summarize the total penguin  contributions in various processes. In our convention, the second meson ($M_2$) corresponds to the one generated from vacuum; and in the lower half of this table, $M_2$ is flavor neutral. The values of Wilson coefficients at the scale $\mu=m_b$ used in our calculation are
$c_1 = 1.105$, $c_2 = -0.228$, $c_3 = 0.013$, $c_4 =-0.029$, $c_5 = 0.009$, $c_6 = -0.033$,
$c_7/\alpha = 0.005$, $c_8/\alpha = 0.060$, $c_9/\alpha= -1.283$, $c_{10}/\alpha = 0.266$ \cite{Buchalla:1995vs}, where $\alpha$ is the fine structure constant.

\begin{table}
  \caption{The penguin contributions to the $B$ meson two-body decays. The meson after the comma, $M_2$, denotes the one produced from vaccum. When  $M_2$ is the flavor neutral meson, the odd coefficients $a_i$ also contribute. They are compiled in the lower half of the table, in addition to the even $a_i$ part. For completeness, we also list the channels involving axial vector mesons.}
\label{tab:penguin}
\begin{ruledtabular}
  \begin{tabular}{lll}
Decay channel &$e_{q'}=+\frac{2}{3}$&$e_{q'}=-\frac{1}{3}$\\
  \hline

$B\rightarrow P_1,P_2$&$a_4+a_{10}+\frac{2m_{P_2}^2}{(m_q+m_{q'})(m_b-m_{q'})}(a_6+a_8)$&$a_4-\frac{1}{2}a_{10}+\frac{2m_{P_2}^2}{(m_q+m_{q'})(m_b-m_{q'})}(a_6-\frac{1}{2}a_8)$\\
$B\rightarrow P,V$&$a_4+a_{10}$&$a_4-\frac{1}{2}a_{10}$\\
$B\rightarrow V,P$&$a_4+a_{10}-\frac{2m_{P}^2}{(m_q+m_{q'})(m_b+m_{q'})}(a_6+a_8)$&$a_4-\frac{1}{2}a_{10}-\frac{2m_{P}^2}{(m_q+m_{q'})(m_b+m_{q'})}(a_6-\frac{1}{2}a_8)$\\
$B\rightarrow V,V$&$a_4+a_{10}$&$a_4-\frac{1}{2}a_{10}$\\
$B\rightarrow A,P$&$a_4+a_{10}+\frac{2m_{P}^2}{(m_q+m_{q'})(m_b-m_{q'})}(a_6+a_8)$&$a_4-\frac{1}{2}a_{10}+\frac{2m_{P}^2}{(m_q+m_{q'})(m_b-m_{q'})}(a_6-\frac{1}{2}a_8)$\\
$B\rightarrow P,A$&$a_4+a_{10}$&$a_4-\frac{1}{2}a_{10}$\\
$B\rightarrow V,A$&$a_4+a_{10}$&$a_4-\frac{1}{2}a_{10}$\\
$B\rightarrow A,V$&$a_4+a_{10}$&$a_4-\frac{1}{2}a_{10}$\\

\hline
$B\rightarrow P_1,P_2^0$&$a_3-a_5+a_{9}-a_7$&$a_3-a_5-\frac{1}{2}(a_{9}-a_7)$\\
$B\rightarrow P,V^0$&$a_3+a_5+a_{9}+a_7$&$a_3+a_5-\frac{1}{2}(a_{9}+a_7)$\\
$B\rightarrow V,P^0$&$a_3-a_5+a_{9}-a_7$&$a_3-a_5-\frac{1}{2}(a_{9}-a_7)$\\
$B\rightarrow V,V^0$&$a_3+a_5+a_{9}+a_7$&$a_3+a_5-\frac{1}{2}(a_{9}+a_7)$\\
$B\rightarrow A,P^0$&$a_3-a_5+a_{9}-a_7$&$a_3-a_5-\frac{1}{2}(a_{9}-a_7)$\\
$B\rightarrow P,A^0$&$a_3-a_5+a_{9}-a_7$&$a_3-a_5-\frac{1}{2}(a_{9}-a_7)$\\
$B\rightarrow V,A^0$&$a_3-a_5+a_{9}-a_7$&$a_3-a_5-\frac{1}{2}(a_{9}-a_7)$\\
$B\rightarrow A,V^0$&$a_3+a_5+a_{9}+a_7$&$a_3+a_5-\frac{1}{2}(a_{9}+a_7)$\\

\end{tabular}
\end{ruledtabular}
\end{table}

From Table \ref{tab:penguin} one can easily read out the amplitude for a given decay process. However, some decay amplitudes contain more than one class. We take $B^+\rightarrow \pi^+\eta$ as an example. In this decay either $\eta$ or $\pi$ can be produced from vacuum. For the case when $\eta$ is produced from vacuum, the decay amplitude can be written as follows.
\begin{equation}
\begin{split}
&q'=u,\, q=d,\, e_{q'}=\frac{2}{3}: ~~~~A_1=\frac{G_F}{\sqrt{2}}\big\{V_{ub}^*V_{ud}a_2-V_{tb}^*V_{td}\big[a_3-a_5+a_{9}-a_7\big]\big\}X_{B^+\to\pi^+,\eta_u}\,,\\
&q'=d,\, q=d,\, e_{q'}=-\frac{1}{3}:  ~~A_2=\frac{G_F}{\sqrt{2}}\big\{-V_{tb}^*V_{td}\left[a_3-a_5-\frac{1}{2}a_{9}+\frac{1}{2}a_7+a_4-\frac{1}{2}a_{10}\right.\\
&~~~~~~~~~~~~~~~~~~~~~~~~~~~~~~~~~~~~~~~~\left.+\frac{m_{\eta}^2}{m_s(m_b-m_d)}\big(a_6-\frac{1}{2}a_8\big)\Big(\frac{f_{\eta}^s}{f_{\eta}^u}-1\Big)
r_{\eta}\right]\Bigr\}X_{B^+\to\pi^+,\eta_u}\,,\\
&q'=s,\, q=d,\, e_{q'}=-\frac{1}{3}: ~~  A_3=\frac{G_F}{\sqrt{2}}\big\{-V_{tb}^*V_{td}\big[a_3-a_5-\frac{1}{2}(a_{9}-a_7)\big]\Bigr\}X_{B^+\to\pi^+,\eta_s}\,,\\
&X_{B^+\to\pi^+,\eta_u}=\langle\pi^+|(\overline{b}d)_{V-A}|B^+\rangle\langle\eta|(\overline{u}u)_{V-A}|0\rangle
=\langle\pi^+|(\overline{b}d)_{V-A}|B^+\rangle f_\eta^u\,,\\
&X_{B^+\to\pi^+,\eta_s}=\langle\pi^+|(\overline{b}d)_{V-A}|B^+\rangle\langle\eta|(\overline{s}s)_{V-A}|0\rangle
=\langle\pi^+|(\overline{b}d)_{V-A}|B^+\rangle f_\eta^s\,.
\end{split}
\end{equation}
The definition of $r_\eta$ is given later in Eq.~\eqref{30}. For the case when $\pi$ is produced from vacuum, the decay amplitude reads
\begin{equation}
\begin{split}
&q'=u,\, q=d,\, e_{q'}=\frac{2}{3}:\\
&A_4=\frac{G_F}{\sqrt{2}}\big\{V_{ub}^*V_{ud}a_1-V_{tb}^*V_{td}\big[a_4+a_{10}+\frac{2m_\pi^2}{(m_u+m_d)(m_b-m_u)}(a_6+a_8)\big]\big\}X_{B^+\rightarrow \eta,\pi^+}\,,\\
&X_{B^+\rightarrow \eta,\pi^+}=\langle\pi|(\overline{u}d)_{V-A}|0\rangle\langle\eta|(\overline{b}u)_{V-A}|B\rangle\,.
\end{split}
\end{equation}
The total amplitude for $B^+\rightarrow \pi^+\eta $ is then given by the sum of these amplitudes
\begin{equation}
A(B\rightarrow \pi^+\eta)=A_1+A_2+A_3+A_4\,.
\end{equation}

\section{The input}
\label{sec:IV}
In our consideration we use the following quark compositions of the light mesons
\begin{equation} \label{1a}
 \begin{split}
   &K^+=u\overline{s}, \, K^0=d\overline{s},\, K^-=s\overline{u}, \\
   &\pi^+(\rho^+)=u\overline{d}, \,\,  \pi^0(\rho^0)=\frac{u\overline{u}-d\overline{d}}{\sqrt{2}}, \,\, \pi^-(\rho^-)=d\overline{u},\\
   &\eta_0=\frac{d\overline{d}+u\overline{u}+s\overline{s}}{\sqrt{3}},\, \eta_8=\frac{d\overline{d}+u\overline{u}-2s\overline{s}}{\sqrt{6}},\\
   &\eta=\eta_8\cos\theta - \eta_0\sin\theta,\, \eta'=\eta_8\sin\theta + \eta_0\cos\theta,
\end{split}
 \end{equation}
with $\theta=-15.4^{\circ}$, which corresponds to the mixing angle $\phi=39.3^\circ$ \cite{Feldmann:1998vh}. Note that such value of $\phi$ was
previously used in Refs.~\cite{Cheng:2017pcq,Cheng:2017qpv}. This value of $\phi$ agrees with the CLEO measurement $42^\circ \pm 2.8^\circ$
\cite{Hietala:2015jqa} and also with the recent BESIII measurement $40.1^\circ \pm 2.1^\circ \pm 0.7^\circ$ \cite{BESIII:2019qci}. Reference \cite{Feldmann:1998vh} presents a nice analysis of the $\eta-\eta'$ mixing both from the theoretical and phenomenological standpoint, where in the former only the masses of pseudoscalar mesons are involved as inputs and in the latter the experimental measurements of branching fractions are used. The relation between the decay constants $f_{\eta^{(')}}^u$ and $f_{\eta^{(')}}^s$ in the singlet-octet mixing scheme and the ones $f_q$ and $f_s$ in the quark flavor basis $q\bar q=\frac{1}{\sqrt{2}}(u\bar u+d\bar d)$ and
$s\bar s$ is given by
\begin{equation}
\begin{split}
f_\eta^u&=\frac{1}{\sqrt{2}}f_q\cos\phi, \qquad f_\eta^s=-f_s\sin\phi,\\
f_{\eta'}^u&=\frac{1}{\sqrt{2}}f_q\sin\phi, \qquad f_\eta^s=f_s\cos\phi,
\end{split}
\end{equation}
with $f_q/f_\pi=1.07, f_s/f_\pi=1.34$.
The values of the decay constants used in our calculations are  \cite{Rosner:2015wva,Ebert:2006hj,Bazavov:2017lyh,Maris:1999nt,Ball:2004rg,Ball:2006eu,Bloch:1999vka} as follows (in MeV)
\begin{equation}
\begin{split}
 &f_\pi=130.2, ~f_K=155.6, ~f_{K^*}=217, \\
 &f_\eta^u=78, ~f_\eta^s=-112, ~f_{\eta'}^u=63, ~f_{\eta'}^s=137, \\
 &f_{D^+}=212.7, ~f_{D^0}=211.6, ~f_{D_s}=249.9, \\
 &f_\rho=205, ~f_\omega=187, ~f_\phi=215.
 \end{split}
\end{equation}

Calculating the matrix elements of the scalar and pseudoscalar currents, one needs to use the equations of motion, Eq.~\eqref{eq:motion}.
When $\eta^{(')}$ is generated from vacuum, the hadron matrix element is treated differently due to the SU(3) breaking \cite{Cheng:1998uy,Ali:1997nh}:
\begin{equation}\label{30}
  \begin{split}
  &\langle\eta^{(')}|\bar s\gamma^5 s|0\rangle=-i\frac{m_{\eta^{(')}}}{2m_s}(f_{\eta^{(')}}^s-f_{\eta^{(')}}^u),\\
  &\langle\eta^{(')}|\bar u\gamma^5 u|0\rangle=\langle\eta^{(')}|\bar d\gamma^5 d|0\rangle=r_{\eta^{(')}}\langle\eta^{(')}|\bar s\gamma^5 s|0\rangle,\\
  &r_{\eta^{'}}=\frac{\sqrt{2f_0^2-f_8^2}}{\sqrt{2f_8^2-f_0^2}}\left(\frac{\cos\theta+\frac{1}{\sqrt{2}}\sin\theta}{\cos\theta-\sqrt{2}\sin\theta}\right),\\
   &r_{\eta}=-\frac{1}{2}\frac{\sqrt{2f_0^2-f_8^2}}{\sqrt{2f_8^2-f_0^2}}\left(\frac{\cos\theta-\sqrt{2}\sin\theta}{\cos\theta+\frac{1}{\sqrt{2}}\sin\theta}\right),
 \end{split}
\end{equation}
with $f_0/f_\pi=1.17$ and $f_8/f_\pi=1.26$. The axial-vector anomaly effect has been incorporated into this equation in order to ensure the correct behavior in the chiral limit. By using Eq.~(2.12) and Eq.~(2.18) from Ref.~\cite{Feldmann:1998vh}, we have
\begin{equation}
\begin{split}
\langle 0|\bar u\gamma^5 u|\eta\rangle&=-\frac{i}{\sqrt{2}}\frac{m_\pi^2}{2m_u}f_q\cos\phi,\\
\langle 0|\bar s\gamma^5 s|\eta\rangle&=-i\frac{2m_K^2-m_\pi^2}{2m_s}f_s\sin\phi.
\end{split}
\end{equation}
Considering the fact that in the chiral limit
\begin{equation}
\frac{m_\pi^2}{2m_u}=\frac{2m_K^2-m_\pi^2}{2m_s},
\end{equation}
we arrive at Eq.~\eqref{30}. Note also that in the limit of $f_0=f_8$, Eq.~\eqref{30} reproduces Eq.~(19) of Ref.~\cite{Akhoury:1987ed}.

The running quark masses at the scale $\mu=m_b$ have the following values \cite{Xing:2007fb}
\begin{equation}
 m_u=1.86,\, m_d=4.22,\, m_c=901,\, m_s=80,\, m_b=4200,
\end{equation}
in units of MeV.
For  the CKM matrix we use  the Wolfenstein parameterization
\begin{equation}
\left(
\begin{array}{ccc}
1-\frac{\lambda^2}{2}&\lambda&A\lambda^3(\rho-i\eta)\\
-\lambda&1-\frac{\lambda^2}{2}&A\lambda^2\\
A\lambda^3(1-\rho-i\eta)&-A\lambda^2&1
\end{array}
\right),
\end{equation}
with central values  $\lambda=0.2265$,\, $A=0.790$,\, $\overline{\rho}=0.141$ and $\overline{\eta}=0.357$ taken from PDG \cite{ParticleDataGroup:2020ssz}.

We employ the form factor values from Refs.~\cite{Faustov:2019mqr} and \cite{Faustov:2022ybm} calculated in RQM, which have been well tested in the semileptonic decays. These form factors are in agreement with lattice determination, and the resulting observables (not only the branching fractions but also the forward-backward asymmetries, polarizations of the leptons or the vector mesons), agree with lattice and experimental results.  As the function of momentum transfer squared, the relevant form factors are expressed by
\begin{itemize}
\item $f_{+}(q^2),\, V(q^2),\, A_{0}(q^2)$:
\begin{equation}
\label{fit:fv}
F(q^2)=\frac{F(0)}{\left(1-\frac{q^2}{M^2}\right)\left[1-\sigma_1\frac{q^2}{M_1^2}+\sigma_2\left(\frac{q^2}{M_1^2}\right)^2\right]}\,,
\end{equation}
\item $f_{0}(q^2),\, A_{1}(q^2),\, A_{2}(q^2)$:
\begin{equation}
\label{fit:a12}
F(q^2)=\frac{F(0)}{1-\sigma_1\frac{q^2}{M_1^2}+\sigma_2\left(\frac{q^2}{M_1^2}\right)^2}\,.
\end{equation}
\end{itemize}
For the $c\rightarrow s$ transition, $M=M_{D_s^*}=2.112$ GeV for the form factors $f_+(q^2),\, V(q^2)$, and $M=M_{D_s}=1.968$ GeV for
the form factor $A_0(q^2)$. For the $c\rightarrow d$ transition, $M=M_{D^*}=2.010$ GeV for the form factors $f_+(q^2),\, V(q^2)$, and $M=M_{D}=1.870$ GeV for the form factor $A_0(q^2)$. For the $b\rightarrow c$ transition, $M=M_{B_c^*}=6.332$ GeV  for the form factors $f_+(q^2),\, V(q^2)$, and $M=M_{B_c}=6.227$ GeV for the form factor $A_0(q^2)$. For the $b\rightarrow u$ transition, $M=M_{B^*}=5.325$ GeV for the form factors $f_+(q^2),V_(q^2)$, and $M=M_{B}=5.280$ GeV for the form factor $A_0(q^2)$. The mass $M_1$ is always taken as the pole mass between the active quarks: $M_1=M_{D_s^*}$ for $c\to s$ transition, $M_1=M_{D^*}$ for $c\to d$ transition, $M_1=M_{B_c^*}$ for $b\to c$ transition, $M_1=M_{B^*}$ for $b\to u$ transition. For convenience, we compile these mass parameters for the charm meson decays in Table \ref{tab:polemassD} while for the bottom meson case one refers to Table \uppercase\expandafter{\romannumeral1} in Ref.~\cite{Faustov:2022ybm}. The values of $F(0),\,\sigma_1,\,\sigma_2$ are easily found in Refs.~\cite{Faustov:2019mqr,Faustov:2022ybm}.

\begin{table}
\caption{Masses in parameterizations of the weak decay form factors of $D$ and $D_s$, cf.~Eqs.~\eqref{fit:fv} and \eqref{fit:a12}.}
\label{tab:polemassD}
\begin{ruledtabular}
  \begin{tabular}{ccccc}
Quark transition  &Decay  &$M_1$ (GeV)  &\multicolumn{2}{c}{$M$ (GeV)}\\
\cline{4-5}
                  &       &             &$f_+(q^2),V(q^2)$    &$A_0(q^2)$\\
  \hline
$c\rightarrow s$   &$D\rightarrow K$    &$2.112$  &$2.112$   &$1.968$\\
                   &$D_s\rightarrow\eta^{(')},\,\phi$  &   &  & \\
  \hline
$c\rightarrow d$   &$D\rightarrow \omega,\pi,\rho,\eta^{(')}$   &$2.010$   &$2.010$    &$1.870$ \\
                   &$D_s\rightarrow K,\phi$                   &     &    &\\
\end{tabular}
\end{ruledtabular}
\end{table}

\section{Results and discussions }
\label{sec:V}
\subsection{Branching fractions}
 The decay branching fractions can be calculated by the equation
\begin{equation}
\mathcal{B}=\tau\frac{|{\bf p}|}{8\pi m^2}|A|^2,\\
\end{equation}
where for the two-body nonleptonic decays, $\tau$ and $m$ are the lifetime and mass of the parent particle, respectively, $|{\bf p}|$ is the magnitude of the three-momentum of the final mesons in the rest frame of the decaying heavy meson, and expressions for the amplitudes $A$ are given in Appendices A and B. The involved expressions for $X$ are given in Eqs.~\eqref{eq:PP}, \eqref{eq:PV}, \eqref{eq:VP}, and \eqref{eq:VV} for $PP,\, PV,\, VP$, and $VV$ modes, respectively. For $VV$ modes, the helicities $H_+,\, H_-$ and $H_0$ are involved. The results for the branching fractions are shown in Tables \ref{res:BrD} and \ref{res:BrB} for charm and bottom meson decays, respectively.

For charm meson decays, both $N=3$ (the number of colors in reality) and the limit $N\to\infty$ are considered in Table \ref{res:BrD}. As mentioned
above, the case of $N\to\infty$ compensates the nonfactorizable effects to some extent and is expected to improve the theoretical predictions empirically. We confirm this point, e.g., for the channels $D^+\to \pi^0\pi^+$, $D^0\to K^+K^-$, $D^0\to K^-\rho^+$ the results for $N\to\infty$ are altered by a factor of about 2 compared to the ones for $N=3$, improving the agreement with the experimental values. For the channels $D^+\to\pi^+\phi$, $D^0\to\eta\eta$, $D_s\to K^+\bar K^0$ and $D_s\to K^+\pi^0$, the effect is even more pronounced, the results are changed by one or two orders of magnitude compared to the ones for $N=3$ bringing them closer to the measured values. In Ref.~\cite{Biswas:2015aaa}, a more elaborate phenomenological analysis is performed, where the annihilation and exchange contributions as well as the resonant final-state interaction (FSI) are considered. As a result, the branching fractions for $D\to KK$ and $D\to \pi\pi$ as a long-standing puzzle get correctly treated in
Ref.~\cite{Biswas:2015aaa} compared to the experimental values. We find in our simple treatment that only the value for the $D^+\to\pi^0\pi^+$ channel agrees with the experimental value within 2 standard deviation while for the $D^0\to\pi^-\pi^+, \pi^0\pi^0$ ones the branching fractions differ from experiment by a factor of $2\sim3$. This is in line with the observation of Ref.~\cite{Biswas:2015aaa} showing the importance of the nonfactorizable effects.

Here we discuss the rule of $N\to\infty$ in more detail. The phenomenon that this rule greatly improves predictions for branching fractions of the nonleptonic two-body $D$ decays was realized by the community in 1980s, as shown in Refs.~\cite{Tadic:1982vn,Bauer:1984zv,Buras:1985xv}. In Ref.~\cite{Buras:1985xv}, Buras and Gerard make a more complete analysis of charm decays, where the effectiveness of $N\to\infty$ is clearly demonstrated compared to the case of $N=3$, and also the result of the $1/N$ expansion is phrased much better in terms of simple diagrammatical rules. But we stress that this rule is purely empirical. As mentioned in Ref.~\cite{Tadic:1982vn}, it is not clear whether this rule is just a coincidence or has a deeper meaning. Note that the generalization of the $N\to\infty$ to the $B$ decays will lead to predictions in contradiction with experiment. Also, the $1/N$ suppression varies in different channels and is rather of a dynamical origin.

In cases where the rule of $N\to\infty$ works well, we can understand what happens for the factorizable and nonfactorizable contributions in the spirit of the large $N$ QCD \cite{Buras:1985xv}. In the usual procedure, the $1/N$ term in Eq.~\eqref{eq:aici}, being part of the factorizable term, is kept while the nonfactorizable term is not considered since there is no reliable way to calculate it. In such a situation the leading and nonleading $1/N$ contributions mix up. The nonfactorizable one, e.g., the final state interaction effect, is nonleading in the $1/N$ expansion. By dropping the $1/N$ term in Eq.~\eqref{eq:aici}, one will work in a self-consistent expansion of $1/N$. Or we can say that the $1/N$ term in factorizable part is almost compensated by the (unknown) nonfactorizable one. There is an explicit calculation to demonstrate this point \cite{Halperin:1994hg}, where the author shows that the soft gluon exchange mechanism (a type of nonfactorizable contribution) tends to cancel the $1/N$ term in the factorized amplitude by using the light cone sum rule. In a more physical picture, we can say that the quarks belonging to different color singlet currents do not easily form a meson and thus the $1/N$ term is highly suppressed.

The results for the $B$ decays are shown in Table \ref{res:BrB}. The theoretical predictions should be better consistent with the experimental
data than in the $D$ meson case. Indeed, the factorization assumption works better for the heavier $B$ meson since the final mesons carry larger momenta. And for some decay channels, such as $B^+\to \rho^+ \eta^{(')}$ and $B_s\to D^{*-}_s \rho^+$, the results for $N=3$ perfectly match the
experimental values within $1.5\,\sigma$ uncertainty. We have calculated branching fractions for the three sets of color number $N$. Those results constitute a range of branching fractions varying with the choices of $N$, which may be understood as an error estimate in some sense. However, there is an exception, for the penguin governed decay $B^+\to\pi^+\phi$  the result for $N=2$ deviates from the one for $N=3$ by two orders of magnitude. We should compare our results to the ones given in Ref.~\cite{Chen:1999nxa} since we work in the same framework, considering the tree-level as well as penguin contributions. However, in Ref.~\cite{Chen:1999nxa} a different set of Wilson coefficients (known as the generalized factorization) is used. Besides, we employ the form factor values predicted by our relativistic quark model, as a more advanced tool from today's perspective compared to their BSW ones. Our results for $N=3$ are of similar magnitude with Ref.~\cite{Chen:1999nxa} under the same condition $N_c^\text{eff}(LL)= N_c^\text{eff}(LR)=3$. For most of channels, our ranges of branching fractions formed by $N=2, 3, \infty$ are close to the corresponding experimental values within $2\,\sigma$ uncertainty.  But there are a few channels where results differ from experimental ones by larger or around factor of
5. Then we also compare with the predictions of Refs.~\cite{Chen:1999nxa,Chen:1998dta}, and find such deviations also happens in their results. In general, for the color-suppressed decay channels (involving $a_2$ terms), such as $B^0\to \pi^0\pi^0$, $B^0\to
\pi^0\rho^0$, $B^0\to \omega\omega$ and $B^0\to \rho^0\rho^0$, the predicted branching fractions are lower than the experiment values. One
of the reasons is due to the smallness of $a_2$, but more importantly, the strong FSI effects should play an essential role, as has been explicitly demonstrated in Ref.~\cite{Chen:1999nxa} in detail. In fact, as we know, the interaction between pseudoscalar octet, e.g., the $\pi\pi-K\bar K$ system, is very strong, for which some of our recent investigations can be found in Refs.~\cite{Wang:2022vga,Dai:2018fmx,Kang:2013jaa,Dai:2017tew,Dai:2014zta}. That is, the $B\to\pi\pi$ decay will receive large contributions from the intermediate states $K\bar K$ and $\eta\eta$ etc. On another hand, it
has been found in Ref.~ \cite{Du:1995my} that the spacelike penguin contributions may be sizable in $B\to PP$ decays, where the authors showed that such corrections to the branching fraction for $B\to\pi\pi$ may be more than 100\%. However, in Ref.~\cite{Ali:1998eb} the authors assume that such contributions in $B\to PV, VV$ decays are not as severe as in $B\to PP$. Reference \cite{Chen:1999nxa} provides a careful examination but those effects of FSI and spacelike penguins can not be reliably determined yet. So conservatively speaking, the branching fraction can be trusted by its order of magnitude.

In this paragraph we give a few comments on comparison of our results with the ones in Ref.~\cite{Chen:1999nxa} by Cheng et al. In this reference many sets of numbers for the branching fraction values are calculated, and these numbers constitute an interval. Such an interval may contain the experimental value, which is very encouraging. But in some cases their ranges span two orders of magnitude. For example, for the $B\to\omega\omega$ decay it reads $7 \times 10^{-8} \sim 2 \times 10^{-6}$, and for the $B\to \rho^0\rho^0$ $5 \times 10^{-8} \sim 2.57 \times 10^{-6}$. Their preferred values correspond to $N_c^\text{eff}(LL)=2$ and $N_c^\text{eff}(LR)=5$. For the same value for the color number, the differences between results of Ref.~\cite{Chen:1999nxa} and ours mainly come from the different inputs. Especially, in Ref.~\cite{Chen:1999nxa} complex-valued numbers for the set of the Wilson coefficients are used while we use the real-valued ones from Buchalla et al in Ref.~\cite{Buchalla:1995vs}. Our main goal is not to reproduce the experimental values exactly or to match them well. We want to test the factorization hypothesis by using our most recent form factor values calculated from an advanced relativistic quark model. To get a more quantitative calculation, the nonfactorizable contribution should be included anyway.

As is known and also mentioned earlier, the nonfactorizable effects may dominate in a specific decay, and there is currently no method to calculate them beforehand. However, in literature there are important works dedicated to the analysis of such nonfactorizable effects by confronting with experimental data. One typical example is the factorization-assisted topological (FAT) approach \cite{Li:2012cfa,Qin:2013tje,Zhou:2015jba,Zhou:2016jkv} which combines the naive factorization hypothesis and the topological diagram approach \cite{Cheng:2010ry,Cheng:2014rfa}. In these papers the authors identify the possible sources of nonfactorizable contributions and then parameterize them, in order to fit  to the existing experimental data. It is found that with the inclusion of the factorization, FAT generally works better than the topological diagram approach, with less parameters and better $\chi^2$ per degree of freedom. Specifically, in Ref.~\cite{Li:2012cfa} for the analysis of $D\to PP$ decays, the authors assign a nonfactorizable term (magnitude and phase) to each of the color-suppressed, $W$-exchange and $W$-annihilation amplitudes, and the Glauber phase is additionally associated to a pion (which is important to resolve the $\pi^+\pi^-$ and $K^+K^-$ branching fraction puzzles). Then 12 parameters are used to fit 28 $D\to PP$ branching fractions and good results are achieved. For the $D\to PV$ decays \cite{Qin:2013tje}, two more parameters are involved compared to the $D\to PP$ ones, with 33 experimental numbers of branching fractions in total. Once the parameters are determined, the authors predict the CP asymmetries for $D$ decays. The results in Ref.~\cite{Zhou:2015jba} are very impressive. The 4 universal parameters are associated to the color-suppressed amplitude and the $W$-exchange amplitude, i.e., parameterizing their sizes and phases, which are used to describe the 31 decay branching fractions induced by the $b\to c$ transition in the $B\to D^{*}M$ decays with $M$ denoting a light pseudoscalar or a vector meson. If available in experiment, the predicted values are consistent with them. Then other 120 decay branching fractions are predicted. The similar analysis of the charmless two-body non-leptonic $B$ decays $B\to PP, PV$ is done in Ref.~\cite{Zhou:2016jkv}. In brief, the nonfactorizable contributions require a fine analysis which is essential for a quantitative prediction of the branching fractions, and it is worth working in this direction in the future.

Here we stress again that we use the most recent form factor values. This is one of our important motivations and improvements. It is known
that the form factors, which encode the underlying dynamics, play a significant role in calculations of the nonleptonic decays, as also
noted in Ref.~\cite{Beneke:2003zv}. In the earlier works \cite{Beneke:2003zv,Chen:1999nxa, Chen:1998dta}, the authors use the form factor values from the sum rule calculations, which are more appropriate for the small values of the momentum-transfer to leptons, or use the older predictions from the BSW model \cite{Chen:1999nxa}. In our case, the RQM includes all sources of relativistic effects, and the form factors are obtained in the whole kinematically allowed region without any extrapolations. Transitions like $B\to D$ and $B\to D^*$ are also considered without using the heavy quark limit.

Moreover, we have calculated as many channels as possible. Previous papers studied only some of them (although the more advanced tools in a formal perspective were used). For example, in Ref.~\cite{Beneke:2000ry} only two decay modes are discussed. In Ref.~\cite{Beneke:2003zv} the $B\to PP$, $PV$ channels are calculated, but not the $B\to VV$ case. We have performed a complete calculation of the $B_{(s)}$, $D_{(s)}$ decay to $PP$, $PV$ and $VV$. In this way, we could show how the form factors influence the results from a holistic point of view based on such framework. So our calculations should be, at least, a useful complement and an important update for the previous ones.

\begin{center}
\LTcapwidth=\linewidth
\begin{longtable}{p{8em}p{7em}p{7em}p{9em}p{9em}}
\caption{Branching fractions of charm meson decays compared to experimental values in PDG \cite{ParticleDataGroup:2020ssz}.
The results for $N=3$ and $N=\infty$ are shown with $N$ being the number of colors. We also list the results corresponding to ``With FSI'' in Ref.~\cite{Biswas:2015aaa}.}
\label{res:BrD}\\ 
  \hline
  \hline
Decay channel&$N=3$&$N\rightarrow \infty$  &Ref.~\cite{Biswas:2015aaa}&PDG \cite{ParticleDataGroup:2020ssz} \\
\hline
\endfirsthead
\multicolumn{5}{c}
{\tablename\ \thetable\ -- \textit{Continued from previous page}} \\
  \hline
  \hline
Decay channel&$N=3$&$N\rightarrow \infty$  &Ref.~\cite{Biswas:2015aaa}&PDG \cite{ParticleDataGroup:2020ssz} \\
\hline
\endhead
\hline \multicolumn{5}{r}{\textit{Continued on next page}} \\
\endfoot
  \hline
  \hline
\endlastfoot

$D^+\to\pi^0\pi^+$&$2.30\times10^{-3}$&$1.30\times10^{-3}$&$(8.89\pm4.51)\times10^{-4}$ &$(1.247\pm0.033)\times10^{-3}$\\
$D^+\to \pi^0 K^+$ &$1.89\times10^{-4}$&$2.52\times10^{-4}$&$(3.07\pm1.02)\times10^{-4}$&$(2.08\pm0.21)\times10^{-4}$ \\
$D^+\to \eta K^+$&$2.25\times10^{-4}$ &$3.01\times10^{-4}$&$(0.98\pm0.26)\times10^{-4}$&$(1.25\pm0.16)\times10^{-4}$ \\
$D^+\to \eta' K^+$&$9.03\times10^{-5}$&$1.21\times10^{-4}$&$(1.40\pm0.39)\times10^{-4}$&$(1.85\pm0.2)\times10^{-4}$\\
$D^+\to \eta\pi^+$&$2.26\times10^{-3}$ &$3.10\times10^{-4}$&$(4.72\pm0.21)\times10^{-3}$&$(3.77\pm0.09)\times10^{-3}$ \\
$D^+\to \eta'\pi^+$&$1.77\times10^{-3}$ &$3.66\times10^{-3}$&$(6.76\pm2.19)\times10^{-3}$&$(4.97\pm0.19)\times10^{-3}$ \\

$D^+\to \pi^+\rho^0$ &$1.43\times10^{-3}$&$2.43\times10^{-4}$&$-$&$(8.3\pm1.5)\times10^{-4}$\\
$D^+\to \pi^+\phi$&$6.93\times10^{-5}$&$2.23\times10^{-3}$&$-$&$(5.7\pm0.14)\times10^{-3}$\\
$D^+\to \pi^+\omega$&$1.16\times10^{-3}$ &$1.91\times10^{-4}$&$-$&$(2.8\pm0.6)\times10^{-4}$ \\
$D^+\to K^+\rho^0$&$1.23\times10^{-4}$&$1.65\times10^{-4}$&$-$&$(1.9\pm0.5)\times10^{-4}$\\
$D^+\to \phi\rho^+$&$8.52\times10^{-5}$ &$2.74\times10^{-3}$&$-$&$<1.5\times10^{-2}$\\

$D^0\to K^-\pi^+$ &$4.07\times10^{-2}$&$5.44\times10^{-2}$&$(3.70\pm1.33)\times10^{-2}$&$(3.950\pm0.031)\times10^{-2}$\\
$D^0\to\pi^-\pi^+$&$2.14\times10^{-3}$&$2.86\times10^{-3}$&$(1.44\pm0.027)\times10^{-3}$&$(1.455\pm0.024)\times10^{-3}$\\
$D^0\to\pi^0\pi^0$&$7.3\times10^{-6}$&$2.35\times10^{-4}$&$(1.14\pm0.56)\times10^{-3}$&$(8.26\pm0.25)\times10^{-4}$\\
$D^0\to K^- K^+$&$2.97\times10^{-3}$&$3.96\times10^{-3}$&$(4.06\pm0.77)\times10^{-3}$&$(4.08\pm0.06)\times10^{-3}$\\
$D^0\to \eta\eta$ &$6.46\times10^{-5}$&$2.07\times10^{-3}$&$(1.27\pm0.27)\times10^{-3}$&$(2.11\pm0.19)\times10^{-3}$\\
$D^0\to \pi^- K^+$&$1.47\times10^{-4}$&$1.97\times10^{-4}$&$(1.77\pm0.88)\times10^{-4}$&$(1.50\pm0.07)\times10^{-4}$\\
$D^0\to \eta\pi^0$&$2.65\times10^{-6}$&$8.50\times10^{-5}$&$(1.47\pm0.90)\times10^{-3}$&$(6.30\pm0.6)\times10^{-4}$\\
$D^0\to \eta'\pi^0$&$6.75\times10^{-6}$&$2.17\times10^{-4}$&$(2.17\pm0.65)\times10^{-3}$&$(9.2\pm0.19)\times10^{-4}$\\
$D^0\to \eta\eta'$&$1.55\times10^{-6}$&$4.98\times10^{-5}$&$(9.53\pm1.83)\times10^{-4}$&$(1.01\pm0.19)\times10^{-3}$\\

$D^0\to \pi^0\omega$ &$1.12\times10^{-6}$&$3.60\times10^{-5}$&$-$&$(1.17\pm0.35)\times10^{-4}$\\
$D^0\to \eta\omega$ &$2.79\times10^{-5}$&$8.95\times10^{-4}$&$-$&$(1.98\pm0.18)\times10^{-3}$\\
$D^0\to \pi^0\rho^0$&$1.91\times10^{-5}$&$6.12\times10^{-4}$&$-$&$(3.86\pm0.23)\times10^{-3}$\\
$D^0\to \pi^-\rho^+$&$4.45\times10^{-3}$&$5.95\times10^{-3}$&$-$&$(1.01\pm0.04)\times10^{-2}$\\
$D^0\to \pi^0\phi$&$1.35\times10^{-5}$ &$4.34\times10^{-4}$&$-$&$(1.17\pm0.04)\times10^{-3}$\\
$D^0\to \rho^-\pi^+$&$1.51\times10^{-3}$&$2.02\times10^{-3}$&$-$&$(5.15\pm0.25)\times10^{-3}$\\
$D^0\to \eta\phi$&$1.20\times10^{-5}$&$3.87\times10^{-4}$&$-$&$(1.8\pm0.5)\times10^{-4}$\\
$D^0\to K^-\rho^+$&$7.94\times10^{-2}$&$1.06\times10^{-1}$&$-$&$(1.13\pm0.07)\times10^{-1}$\\
$D^0\to \eta\bar K^{*0}$&$3.07\times10^{-4}$&$9.86\times10^{-3}$ &$-$&$-$\\
$D^0\to \eta'\bar K^{*0}$&$2.41\times10^{-6}$&$7.72\times10^{-5}$&$-$&$<1.0\times10^{-3}$\\

$D^0\to \rho^0\rho^0$&$2.16\times10^{-5}$& $6.90\times10^{-4}$&$-$&$(1.85\pm0.13)\times10^{-3}$\\
$D^0\to \phi\omega$&$1.46\times10^{-5}$ &$4.68\times10^{-4}$&$-$&$<2.1\times10^{-3}$\\

$D_s\to K^+\bar K^0 $&$4.89\times10^{-4}$&$1.57\times10^{-2}$&$-$&$(2.95\pm0.14)\times10^{-2}$\\
$D_s\to \eta\pi^+$&$2.19\times10^{-2}$&$2.92\times10^{-2}$&$(2.26\pm0.82)\times10^{-2}$&$(1.68\pm0.10)\times10^{-2}$\\
$D_s\to K^+\pi^0$ &$9.83\times10^{-6}$&$3.16\times10^{-4}$&$(8.17\pm4.64)\times10^{-4}$&$(6.21\pm2.1)\times10^{-4}$\\
$D_s\to \eta'\pi^+$ &$1.96\times10^{-2}$&$2.62\times10^{-2}$&$(2.64\pm0.78)\times10^{-2}$&$(3.94\pm0.25)\times10^{-2}$\\
$D_s\to \eta K^+$ &$1.76\times10^{-3}$&$3.97\times10^{-3}$&$(1.50\pm0.75)\times10^{-3}$&$(1.72\pm0.34)\times10^{-3}$\\
$D_s\to \eta' K^+$ &$9.76\times10^{-4}$&$2.99\times10^{-4}$&$(7.07\pm0.49)\times10^{-4}$&$(1.7\pm0.5)\times10^{-3}$\\

$D_s\to \eta\rho^+ $&$4.64\times10^{-2}$&$6.20\times10^{-2}$&$-$&$(8.9\pm0.8)\times10^{-2}$\\
$D_s\to \eta'\rho^+$&$2.09\times10^{-2}$&$2.79\times10^{-2}$ &$-$&$(5.8\pm1.5)\times10^{-2}$\\
$D_s\to K^+\omega$&$2.36\times10^{-5}$&$7.59\times10^{-4}$ &$-$&$(8.7\pm2.5)\times10^{-4}$\\
$D_s\to K^+\rho^0$&$2.83\times10^{-5}$&$9.09\times10^{-4}$ &$-$&$(2.5\pm0.4)\times10^{-3}$\\
$D_s\to \phi\pi^+$&$2.79\times10^{-2}$&$3.73\times10^{-2}$ &$-$&$(4.5\pm0.4)\times10^{-2}$\\
$D_s\to \phi\rho^+$ &$9.92\times10^{-2}$ &$1.33\times10^{-1}$ &$-$&$(8.4^{+1.9}_{-2.3})\times10^{-2}$\\
\end{longtable}
\end{center}
\begin{center}
\LTcapwidth=\linewidth
\begin{longtable}{p{8em}p{8em}<{\centering}p{6em}<{\centering}p{6em}<{\centering}p{11em}<{\centering}p{11em}}
\caption{Branching fractions of bottom meson decays. The results for $N=2, 3$ and $N=\infty$ (with $N$ being the number of colors)
are shown compared to experimental values in PDG \cite{ParticleDataGroup:2020ssz}. We also show the results of Refs.~\cite{Chen:1999nxa,Chen:1998dta} for part of channels for which our theoretical values for $N=3$ deviate experimental ones by larger or around factor of 4.}
\label{res:BrB}\\
  \hline
  \hline
Decay channel  &$N=2$   &$N=3$   &$N=\infty$ &Others  &PDG \cite{ParticleDataGroup:2020ssz} \\
\hline
\endfirsthead
\multicolumn{6}{c}
{\tablename\ \thetable\ -- \textit{Continued from previous page}} \\
\hline \hline
Decay channel&$N=2$&$N=3$&$N=\infty$ &Others &PDG \cite{ParticleDataGroup:2020ssz} \\
\hline
\endhead
\hline \multicolumn{6}{r}{\textit{Continued on next page}} \\
\endfoot
\hline\hline
\endlastfoot

$B^+\to \pi^+\eta$&$5.80\times10^{-6}$&$5.06\times10^{-6}$ &$3.79\times10^{-6}$&$-$ &$(4.02\pm0.27)\times10^{-6}$\\
$B^+\to \pi^+\eta'$&$4.71\times10^{-6}$&$4.19\times10^{-6}$ &$3.32\times10^{-6}$ &$-$ &$(2.7\pm0.9)\times10^{-6}$ \\

$B^+\to \omega\pi^+$&$5.84\times10^{-6}$ &$4.28\times10^{-6}$&$1.92\times10^{-6}$&$-$ &$(6.9\pm0.5)\times10^{-6}$\\
$B^+\to \rho^+\eta$&$1.20\times10^{-5}$ &$1.11\times10^{-5}$&$9.31\times10^{-6}$&$-$ &$(7.0\pm2.9)\times10^{-6}$\\
$B^+\to \rho^+\eta'$&$1.02\times10^{-5}$ &$9.58\times10^{-6}$&$8.43\times10^{-6}$&$-$ &$(9.7\pm2.2)\times10^{-6}$\\

$B^+\to \pi^+ K^0$&$3.36\times10^{-6}$&$3.94\times10^{-6}$&$5.23\times10^{-6}$&$1.89\times10^{-5}$ \cite{Chen:1999nxa}&$(2.37\pm0.08)\times10^{-5}$\\
$B^+\to \rho^+ K^0$&$3.35\times10^{-7}$&$3.03\times10^{-7}$&$2.45\times10^{-7}$&$2.4\times10^{-7}$ \cite{Chen:1999nxa}&$(7.3^{+1.0}_{-1.2})\times10^{-6}$\\

$B^+\to \pi^0\pi^+$&$4.24\times10^{-6}$&$3.35\times10^{-6}$&$1.89\times10^{-6}$&$-$ &$(5.5\pm0.4)\times10^{-6}$\\
$B^+\to \pi^+\rho^0$&$5.78\times10^{-6}$ &$4.06\times10^{-6}$&$1.55\times10^{-6}$&$-$&$(8.3\pm1.2)\times10^{-6}$\\
$B^+\to \pi^0\rho^+$ &$8.77\times10^{-6}$ &$7.61\times10^{-6}$&$5.53\times10^{-6}$&$-$&$(1.09\pm0.14)\times10^{-5}$\\
$B^+\to \pi^+\phi$ &$2.52\times10^{-11}$ &$2.44\times10^{-9}$&$5.52\times10^{-8}$&$-$&$(3.2\pm1.5)\times10^{-8}$\\
$B^+\to \rho^+\rho^0$ &$1.28\times10^{-5}$&$1.01\times10^{-5}$&$6.67\times10^{-6}$&$-$ &$(2.4\pm0.19)\times10^{-5}$\\
$B^+\to \rho^+\omega$&$1.10\times10^{-5}$&$9.64\times10^{-6}$&$7.10\times10^{-6}$&$-$ &$(1.59\pm0.21)\times10^{-5}$\\

$B^0\to D^-\pi^+$&$3.86\times10^{-3}$&$4.17\times10^{-3}$&$4.80\times10^{-3}$&$-$&$(2.52\pm0.13)\times10^{-3}$\\
$B^0\to D^- K^+$&$2.97\times10^{-4}$&$3.20\times10^{-4}$&$3.69\times10^{-4}$&$-$  &$(1.86\pm0.2)\times10^{-4}$\\

$B^0\to \pi^- K^+$&$3.10\times10^{-6}$ &$3.37\times10^{-6}$&$3.93\times10^{-6}$&$1.56\times10^{-5}$ \cite{Chen:1999nxa}&$(1.96\pm0.05)\times10^{-5}$ \\
$B^0\to \pi^-\pi^+$&$4.54\times10^{-6}$&$4.90\times10^{-6}$&$5.65\times10^{-6}$&$-$ &$(5.12\pm0.19)\times10^{-6}$ \\
$B^0\to\pi^0\pi^0$&$2.62\times10^{-7}$&$7.00\times10^{-8}$&$1.48\times10^{-7}$&$9\times10^{-7}$ \cite{Chen:1999nxa}&$(1.59\pm0.26)\times10^{-6}$\\
$B^0\to \pi^0\eta$&$1.02\times10^{-7}$&$9.60\times10^{-8}$&$1.33\times10^{-7}$&$-$&$(4.1\pm1.7)\times10^{-7}$ \\
$B^0\to \pi^0\eta'$&$7.57\times10^{-8}$ &$4.48\times10^{-8}$&$7.41\times10^{-8}$&$7\times10^{-8}$ \cite{Chen:1999nxa}&$(1.2\pm0.6)\times10^{-6}$ \\
$B^0\to \eta\eta$&$6.80\times10^{-7}$ &$2.69\times10^{-7}$ &$5.13\times10^{-7}$&$-$&$<1.0\times10^{-6}$\\
$B^0\to \eta'\eta'$&$3.58\times10^{-7}$ &$8.50\times10^{-8}$&$2.14\times10^{-7}$&$-$ &$<1.7\times10^{-6}$ \\

$B^0\to D^{*-} \pi^{+}$&$4.93\times10^{-3}$ &$5.32\times10^{-3}$&$6.13\times10^{-3}$&$-$&$(2.74\pm0.13)\times10^{-3}$ \\
$B^0\to D^-\rho^+$&$9.31\times10^{-3}$&$1.00\times10^{-2}$&$1.16\times10^{-2}$&$-$ &$(7.6\pm1.2)\times10^{-3}$ \\
$B^0\to \pi^0\rho^0$ &$8.09\times10^{-7}$ &$1.51\times10^{-7}$ &$4.00\times10^{-7}$&$3\times10^{-8}$ \cite{Chen:1999nxa}&$(2.0\pm0.5)\times10^{-6}$\\
$B^0\to \pi^0\omega$&$1.72\times10^{-8}$ &$3.96\times10^{-9}$&$1.39\times10^{-8}$&$-$&$<5\times10^{-7}$\\
$B^0\to \rho^- K^+$&$7.56\times10^{-7}$ &$8.56\times10^{-7}$ &$1.08\times10^{-6}$ &$1.16\times10^{-6}$  \cite{Chen:1999nxa}&$(7.0\pm0.9)\times10^{-6}$\\

$B^0\to \pi^- K^{*+}$ &$1.10\times10^{-6}$ &$1.17\times10^{-6}$ &$1.30\times10^{-6}$&$6.84\times10^{-6}$ \cite{Chen:1999nxa}&$(7.5\pm0.4)\times10^{-6}$ \\
$B^0\to \rho^-\pi^{+}$&$4.77\times10^{-6}$ &$5.15\times10^{-6}$&$5.94\times10^{-6}$&$8.06\times10^{-6}$\cite{Chen:1999nxa}&$(2.30\pm0.223)\times10^{-5}$ \\
$B^0\to D^- K^{*+}$&$5.58\times10^{-4}$&$6.01\times10^{-4}$&$6.93\times10^{-4}$&$-$&$(4.5\pm0.7)\times10^{-4}$ \\
$B^0\to D^{*-}K^{+}$&$3.73\times10^{-4}$ &$4.02\times10^{-4}$&$4.64\times10^{-4}$&$-$&$(2.12\pm0.15)\times10^{-4}$ \\
$B^0\to \eta\eta'$ &$4.91\times10^{-7}$&$1.54\times10^{-7}$&$3.32\times10^{-7}$&$-$ &$<1.2\times10^{-6}$ \\
$B^0\to \eta\rho^0$&$1.38\times10^{-7}$&$2.65\times10^{-8}$&$6.99\times10^{-8}$&$-$ &$<1.5\times10^{-6}$ \\
$B^0\to \eta'\rho^0$&$1.49\times10^{-7}$ &$2.87\times10^{-8}$&$7.35\times10^{-8}$&$-$&$<1.3\times10^{-6}$ \\
$B^0\to \eta\omega$ &$8.79\times10^{-7}$ &$1.68\times10^{-7}$&$4.32\times10^{-7}$&$-$ &$(9.4^{+4.0}_{-3.1})\times10^{-7}$\\
$B^0\to \eta'\omega$&$7.01\times10^{-7}$ &$1.30\times10^{-7}$&$3.55\times10^{-7}$&$-$ &$(1.0^{+0.5}_{-0.4})\times10^{-6}$ \\
$B^0\to \rho^+\pi^{-}$&$1.10\times10^{-5}$&$1.19\times10^{-5}$&$1.37\times10^{-5}$&$-$ &$(2.30\pm0.223)\times10^{-5}$ \\

$B^0\to \rho^+\rho^-$&$1.34\times10^{-5}$ &$1.45\times10^{-5}$  &$1.67\times10^{-5}$ &$-$&$(2.77\pm0.19)\times10^{-5}$\\
$B^0\to \rho^0\rho^0$&$7.25\times10^{-7}$ &$1.81\times10^{-7}$ &$3.63\times10^{-8}$&$5\times10^{-8}$  \cite{Chen:1999nxa}&$(9.6\pm1.5)\times10^{-7}$\\
$B^0\to \omega\omega$&$3.27\times10^{-7}$ &$6.85\times10^{-8}$ &$1.56\times10^{-8}$ &$7\times10^{-8}$ \cite{Chen:1999nxa}&$(1.2\pm0.4)\times10^{-6}$\\
$B^0\to \omega\rho^0$&$7.11\times10^{-8}$&$2.62\times10^{-8}$ &$2.71\times10^{-8}$&$-$ &$<1.6\times10^{-6}$\\
$B^0\to D^{*-}\rho^{+}$&$1.34\times10^{-2}$ &$1.45\times10^{-2}$ &$1.67\times10^{-2}$ &$-$&$(6.8\pm0.9)\times10^{-3}$\\
$B^0\to D^{*-}K^{*+}$&$8.34\times10^{-4}$ &$8.99\times10^{-4}$ &$1.04\times10^{-3}$ &$-$&$(3.3\pm0.6)\times10^{-4}$\\

$B_s\to D^-_s\pi^+$&$3.56\times10^{-3}$&$3.84\times10^{-3}$&$4.43\times10^{-3}$&$-$ &$(3.00\pm0.23)\times10^{-3}$ \\
$B_s\to D^-_s K^+$&$2.75\times10^{-4}$&$2.96\times10^{-4}$&$3.41\times10^{-4}$&$-$  &$(2.27\pm0.19)\times10^{-4}$ \\

$B_s\to K^-\pi^+$&$7.96\times10^{-6}$&$8.59\times10^{-6}$&$9.91\times10^{-6}$&$-$ &$(5.8\pm0.7)\times10^{-6}$ \\
$B_s\to K^- K^+$&$5.47\times10^{-6}$&$5.94\times10^{-6}$&$6.95\times10^{-6}$&$1.09\times10^{-5}$ \cite{Chen:1998dta} &$(2.66\pm0.22)\times10^{-5}$ \\
$B_s\to D^-_s\rho^+$&$8.60\times10^{-3}$&$9.27\times10^{-3}$ &$1.07\times10^{-2}$&$-$  &$(6.9\pm1.4)\times10^{-3}$ \\
$B_s\to D^{*-}_s\pi^+$&$2.94\times10^{-3}$&$3.17\times10^{-3}$ &$3.66\times10^{-3}$ &$-$ &$(2.0\pm0.5)\times10^{-3}$ \\
$B_s\to D^{*-}_s K^+$&$2.22\times10^{-4}$&$2.39\times10^{-4}$&$2.76\times10^{-4}$ &$-$ &$(1.33\pm0.35)\times10^{-4}$ \\
$B_s\to D^{-}_s K^{*+}$&$5.16\times10^{-4}$&$5.56\times10^{-4}$&$6.41\times10^{-4}$&$-$ &$-$ \\

$B_s\to K^{*-}\pi^+$&$7.67\times10^{-6}$&$8.27\times10^{-6}$&$9.54\times10^{-6}$&$-$ &$(2.9\pm1.1)\times10^{-6}$ \\
$B_s\to K^{*-} K^+$&$1.20\times10^{-6}$&$1.35\times10^{-6}$&$1.70\times10^{-6}$&$7.5\times10^{-7}$ \cite{Chen:1998dta}&$(1.9\pm0.5)\times10^{-5}$ \\

$B_s\to K^{-}K^{*+}$&$1.86\times10^{-6}$&$1.97\times10^{-6}$&$2.20\times10^{-6}$&$3.77\times10^{-6}$ \cite{Chen:1998dta} &$(1.9\pm0.5)\times10^{-5}$ \\
$B_s\to D^{*-}_s K^{*+}$&$5.48\times10^{-4}$&$5.90\times10^{-4}$&$6.81\times10^{-4}$&$-$ &$(1.33\pm0.35)\times10^{-4}$ \\
$B_s\to D_s^{*-}\rho^+$&$8.64\times10^{-3}$&$9.31\times10^{-3}$ &$1.07\times10^{-2}$&$-$  &$(9.6\pm2.1)\times10^{-3}$
\end{longtable}
\end{center}

\subsection{A note on the conventions for the definitions of form factors and decay constants}
In some references \cite{Amsler,QMPDG}, the following quark compositions for the octet mesons are used
\begin{equation}
\begin{split}
&K^+=u\overline{s},\, K^0=d\overline{s},\,  K^-=-s\overline{u},\\
&\pi^+(\rho^+)=u\overline{d},\, \pi^0(\rho^0)=\frac{d\overline{d}-u\overline{u}}{\sqrt{2}},\,  \pi^-(\rho^-)=-d\overline{u}~,\\
\end{split}
\end{equation}
which are different from Eq.~\eqref{1a} for the $K^-, \pi^0, \rho^0, \pi^-, \rho^-$ cases. Then the definitions of the decay constants as well as the corresponding transition form factors will  change by an overall sign. Any physical result is not affected.

We have also checked different conventions on definitions of the form factors and decay constants, which differ by factors of $(-1)$ and/or $i$. Note that this detail may influence the calculation if using an inappropriate/incosistent convention. In the factorization scheme we are treating the product of $\langle M_1|J^\mu|B\rangle \langle M_2|J_\mu|0\rangle$, and surely an overall sign does not matter. However, for e.g., the channel $B\to\pi \rho$ has the subprocess $\langle \pi|J^\mu|B\rangle \langle\rho|J_\mu|0\rangle$ and $\langle \rho|J^\mu|B\rangle \langle\pi|J_\mu|0\rangle$ and thus their interference occurs.  Note also the convention difference $\epsilon^{0123}\equiv -1$ and $\epsilon^{0123}\equiv+1$, where the former is used in Refs.~\cite{Cheng:2003sm, Beneke:2000ry} and the latter is used in Refs.~\cite{Ebert:1997mg,Jaus:1999zv}. We have checked that the final results in Refs.~\cite{Ali:1998eb,Chen:1999nxa,Beneke:2000ry,Beneke:2003zv,Ali:1997nh} agree with each other just up to an overall factor of $(-i), i$ or $-1$, which have no influence for branching fraction of a two-body decay. As a result, the vector decay constant should be defined by \footnote{By taking the hermitian conjugate, one finds $\langle P(q)|\bar q\gamma^\mu\gamma^5 q'|0\rangle=-i f_P q^\mu$ is equivalent to  $\langle 0|\bar q'\gamma^\mu\gamma^5 q|P(q)\rangle=-if_P q^\mu$ and $\langle 0|V_\mu|V(\epsilon, q)\rangle=-if_V m_V \epsilon_\mu$ is equivalent to $\langle V(\epsilon, q)|V_\mu|0\rangle=if_V m_V \epsilon^*_\mu$.}
$\langle V(\epsilon, q)|V_\mu|0\rangle=-if_V m_V \epsilon^*_\mu$ in Ref.~\cite{Ebert:1997mg}.
Then the amplitude for $\bar B^0\to \pi^0\rho^0$ follows:
 \begin{eqnarray}
 A(\bar B^0\to \pi^0\rho^0)=\frac{G_F}{\sqrt{2}}\Big[-i f_\rho m_\rho\epsilon^*\cdot p_\pi F_1(m_\rho^2) -i f_\pi m_\rho\epsilon^*\cdot p_\pi A_0(m_\pi^2)\Big]\,.
\end{eqnarray}
It is also important to mention that all authors use real and positive form-factor and decay-constant values.

\section{Conclusions}
\label{sec:VI}
Based on the form factors computed in the relativistic quark model, we calculate the branching fraction of 100 nonleptonic decay channels of charm and bottom mesons. We provide the detailed derivation for the decay amplitudes and branching fractions. The numerical results are shown in the Sec.~\ref{sec:V}.

For the nonleptonic decay process of the $D$ mesons, we consider only the tree-level contributions and use the number of colors $N=3$ and $N\to\infty$ for demonstration. Taking the value of $N$ different from 3 is a way to parametrize the nonfactorizable effects. And indeed we find
that the limit $N\to\infty$ works much better than other numbers of $N$ generally. Some typical examples are $D^+\to\pi^0\pi^+$, $D^0\to K^-K^+$, $D^0\to\eta\eta$, $D^0\to K^-\rho^+$, and $D_s\to K^+\omega$.

For the nonleptonic decay process of the bottom mesons, we consider both the tree-level and penguin contributions. The results for branching fractions are in agreement with the experimental data for most of decay channels. However, for some decays, e.g., $B^+\to\pi^+ K^0, \rho^+K^0$ and $B^0\to \pi^0\eta', \pi^0\pi^0, \rho^-K^+$ our results are too small, and as it has been demonstrated in Ref.~\cite{Chen:1999nxa}, the final-state interaction effects play an indispensable role to get the quantitatively correct values.

\acknowledgments
We are grateful to Prof.~Hai-Yang Cheng for very valuable discussions on his paper \cite{Chen:1999nxa} and answering our questions quickly. V.O.G. thanks Profs. Dietmar Ebert and Rudolf Faustov for usefull discussions. The author X.W.K. acknowledges the support from the National Natural Science Foundation of China (NSFC) under Project Nos. 11805012 and 12275023.

\appendix
\section{Decay amplitudes of the charm mesons}
Here we list the amplitudes for the charm meson decays. The definition of $X$ is given in Sec.~\ref{sec:II}.
\begin{equation}
  A(D^+\rightarrow \pi^0 \pi^+)=\frac{G_F}{\sqrt{2}}V_{cd}^*V_{ud}\Big[a_1X_{D^+\rightarrow\pi^0,\pi^+}+a_2X_{D^+\rightarrow\pi^+,\pi^0}\Big],
\end{equation}
\begin{equation}
  A(D^+\rightarrow \pi^0 K^+)=\frac{G_F}{\sqrt{2}}V_{cd}^*V_{us}a_1X_{D^+\rightarrow\pi^0,K^+},
\end{equation}
\begin{equation}
    A(D^+\rightarrow \eta^{(')} K^+)=\frac{G_F}{\sqrt{2}}V_{cd}^*V_{us}a_1X_{D^+\rightarrow\eta^{(')},K^+},
\end{equation}
\begin{eqnarray}
  A(D^+\rightarrow\eta^{(')} \pi^+)&=&\frac{G_F}{\sqrt{2}}\Big[V_{cd}^*V_{ud}a_1X_{D^+\rightarrow\eta^{(')},\pi^+}+V_{cd}^*V_{ud}a_2X_{D^+\rightarrow\pi^+,\eta^{(')}_u}\nl
  &&+V_{cs}^*V_{us}a_2X_{D^+\rightarrow\pi^+,\eta^{(')}_s}\Big],
\end{eqnarray}
\begin{equation}
  A(D^+\rightarrow \pi^+\rho^0)=\frac{G_F}{\sqrt{2}}V_{cd}^*V_{ud}\Big[a_1X_{D^+\rightarrow\rho^0,\pi^+}+a_2X_{D^+\rightarrow\pi^+,\rho^0}\Big],
\end{equation}
\begin{equation}
  A(D^+\rightarrow \pi^+ \phi)=\frac{G_F}{\sqrt{2}}V_{cs}^*V_{us}a_2X_{D^+\rightarrow\pi^+,\phi},
\end{equation}
\begin{equation}
  A(D^+\rightarrow \pi^+\omega)=\frac{G_F}{\sqrt{2}}V_{cd}^*V_{ud}\Big[a_1X_{D^+\rightarrow\omega,\pi^+}+a_2X_{D^+\rightarrow\pi^+,\omega}\Big],
\end{equation}
\begin{equation}
  A(D^+\rightarrow \rho^0K^+)=\frac{G_F}{\sqrt{2}}V_{cd}^*V_{us}a_1X_{D^+\rightarrow\rho^0,K^+},
\end{equation}
\begin{equation}
  A(D^+\rightarrow\rho^+ \phi)=\frac{G_F}{\sqrt{2}}V_{cs}^*V_{us}a_2X_{D^+\rightarrow\rho^+,\phi},
\end{equation}
\begin{equation}
  A(D^0\rightarrow K^-\pi^+)=\frac{G_F}{\sqrt{2}}V_{cs}^*V_{ud}a_1X_{D^0\rightarrow K^-,\pi^+},
\end{equation}
\begin{equation}
  A(D^0\rightarrow\pi^- \pi^+)=\frac{G_F}{\sqrt{2}}V_{cd}^*V_{ud}a_1X_{D^0\rightarrow \pi^-,\pi^+},
\end{equation}
\begin{equation}
  A(D^0\rightarrow\pi^0 \pi^0)=2\frac{G_F}{\sqrt{2}}V_{cd}^*V_{ud}a_2X_{D^0\rightarrow \pi^0,\pi^0},
\end{equation}
\begin{equation}
  A(D^0\rightarrow K^- K^+)=\frac{G_F}{\sqrt{2}}V_{cs}^*V_{us}a_1X_{D^0\rightarrow K^-,K^+},
\end{equation}
\begin{equation}
  A(D^0\rightarrow \eta\eta)=2\frac{G_F}{\sqrt{2}}\Big[V_{cs}^*V_{us}a_2X_{D^0\rightarrow\eta,\eta_s}+V_{cd}^*V_{ud}a_2X_{D^0\rightarrow \eta,\eta_u}\Big],
\end{equation}
\begin{equation}
  A(D^0\rightarrow \pi^-K^+)=\frac{G_F}{\sqrt{2}}V_{cd}^*V_{us}a_1X_{D^0\rightarrow \pi^-,K^+},
\end{equation}
\begin{eqnarray}
  A(D^0\rightarrow\eta^{(')} \pi^0)&=&\frac{G_F}{\sqrt{2}}\Big[V_{cd}^*V_{ud}a_2X_{D^0\rightarrow\eta^{(')},\pi^0}+V_{cd}^*V_{ud}a_2X_{D^0\rightarrow\pi^0,\eta_u^{(')}}\nl
 && +V_{cs}^*V_{us}a_2X_{D^0\rightarrow\pi^0,\eta_s^{(')}}\Big],
\end{eqnarray}
\begin{eqnarray}
A(D^0\rightarrow \eta\eta')&=&\frac{G_F}{\sqrt{2}}\Big[V_{cd}^*V_{ud}a_2X_{D^0\rightarrow \eta,\eta'_u}+V_{cs}^*V_{us}a_2X_{D^0\rightarrow\eta,\eta'_s}\nonumber\\
&&+V_{cd}^*V_{ud}a_2X_{D^0\rightarrow \eta',\eta_u}+V_{cs}^*V_{us}a_2X_{D^0\rightarrow\eta',\eta_s}\Big],
\end{eqnarray}
\begin{equation}
  A(D^0\rightarrow \pi^0\omega)=\frac{G_F}{\sqrt{2}}\Big[V_{cd}^*V_{ud}a_2X_{D^0\rightarrow \pi^0,\omega}+V_{cd}^*V_{ud}a_2X_{D^0\rightarrow \omega,\pi^0 }\Big],
\end{equation}
\begin{equation}
  A(D^0\rightarrow \eta\omega)=\frac{G_F}{\sqrt{2}}\Big[V_{cd}^*V_{ud}a_2X_{D^0\rightarrow \eta,\omega}+V_{cd}^*V_{ud}a_2X_{D^0\rightarrow \omega,\eta_u}+V_{cs}^*V_{us}a_2X_{D^0\rightarrow \omega,\eta_s}\Big],
\end{equation}
\begin{equation}
  A(D^0\rightarrow\rho^0\pi^0)=\frac{G_F}{\sqrt{2}}\Big[V_{cd}^*V_{ud}a_2X_{D^0\rightarrow \pi^0,\rho^0}+V_{cd}^*V_{ud}a_2X_{D^0\rightarrow \rho^0,\pi^0}\Big],
\end{equation}
\begin{equation}
  A(D^0\rightarrow\pi^-\rho^+)=\frac{G_F}{\sqrt{2}}V_{cd}^*V_{ud}a_1X_{D^0\rightarrow \pi^-,\rho^+},
\end{equation}
\begin{equation}
  A(D^0\rightarrow\pi^0\phi)=\frac{G_F}{\sqrt{2}}V_{cs}^*V_{us}a_2X_{D^0\rightarrow \pi^0,\phi},
\end{equation}
\begin{equation}
  A(D^0\rightarrow\rho^-\pi^+)=\frac{G_F}{\sqrt{2}}V_{cd}^*V_{ud}a_1X_{D^0\rightarrow \rho^-,\pi^+},
\end{equation}
\begin{equation}
  A(D^0\rightarrow\eta\phi)=\frac{G_F}{\sqrt{2}}V_{cs}^*V_{us}a_2X_{D^0\rightarrow \eta,\phi},
\end{equation}
\begin{equation}
  A(D^0\rightarrow K^-\rho^+)=\frac{G_F}{\sqrt{2}}V_{cs}^*V_{ud}a_1X_{D^0\rightarrow K^-,\rho^+},
\end{equation}
\begin{equation}
  A(D^0\rightarrow\eta^{(')}\bar{K}^{*0})=\frac{G_F}{\sqrt{2}}V_{cs}^*V_{ud}a_2X_{D^0\rightarrow\eta^{(')},\bar{K}^{*0}},
\end{equation}
\begin{equation}
  A(D^0\rightarrow \rho^0\rho^0)=2\frac{G_F}{\sqrt{2}}V_{cd}^*V_{ud}a_2X_{D^0\rightarrow \rho^0,\rho^0},
\end{equation}
\begin{equation}
  A(D^0\rightarrow \omega\phi)=\frac{G_F}{\sqrt{2}}V_{cs}^*V_{us}a_2X_{D^0\rightarrow \omega,\phi},
\end{equation}
\begin{equation}
  A(D_s\rightarrow K^+\bar{K}^0)=\frac{G_F}{\sqrt{2}}V_{cs}^*V_{ud}a_2X_{D_s\rightarrow K^+,\bar{K}^0},
\end{equation}
\begin{equation}
  A(D_s\rightarrow \eta^{(')}\pi^+)=\frac{G_F}{\sqrt{2}}V_{cs}^*V_{ud}a_1X_{D_s\rightarrow \eta^{(')},\pi^+},
\end{equation}
\begin{equation}
  A(D_s\rightarrow K^+\pi^0)=\frac{G_F}{\sqrt{2}}V_{cd}^*V_{ud}a_2X_{D_s\rightarrow K^+,\pi^0},
\end{equation}
\begin{eqnarray}
  A(D_s\rightarrow \eta^{(')}K^+)&=&\frac{G_F}{\sqrt{2}}\Big[V_{cd}^*V_{ud}a_2X_{D_s\rightarrow K^+,\eta^{(')}_u}+V_{cs}^*V_{us}a_2X_{D_s\rightarrow K^+,\eta^{(')}_s}\nonumber\\
  &&+V_{cs}^*V_{us}a_1X_{D_s\rightarrow \eta^{(')},K^+}\Big],
\end{eqnarray}
\begin{equation}
  A(D_s\rightarrow \eta^{(')}\rho^+)=\frac{G_F}{\sqrt{2}}V_{cs}^*V_{ud}a_1X_{D_s\rightarrow \eta^{(')},\rho^+},
\end{equation}
\begin{equation}
  A(D_s\rightarrow K^+\omega)=\frac{G_F}{\sqrt{2}}V_{cd}^*V_{ud}a_2X_{D_s\rightarrow K^+,\omega},
\end{equation}
\begin{equation}
  A(D_s\rightarrow K^+\rho^0)=\frac{G_F}{\sqrt{2}}V_{cd}^*V_{ud}a_2X_{D_s\rightarrow K^+,\rho^0},
\end{equation}
\begin{equation}
  A(D_s\rightarrow \phi\pi^+)=\frac{G_F}{\sqrt{2}}V_{cs}^*V_{ud}a_1X_{D_s\rightarrow \phi,\pi^+},
\end{equation}
\begin{equation}
  A(D_s\rightarrow \phi\rho^+)=\frac{G_F}{\sqrt{2}}V_{cs}^*V_{ud}a_1X_{D_s\rightarrow \phi,\rho^+}.
\end{equation}

\section{Decay amplitudes of the bottom mesons}

As in Appendix A, here the amplitude for the bottom meson decays are provided.
\begin{eqnarray}
  A(B^+\rightarrow \pi^+\eta^{(')})&=&\frac{G_F}{\sqrt{2}}\left\{\left[V_{ub}^*V_{ud}a_2-V_{tb}^*V_{td}\left(2a_3-2a_5-\frac{1}{2}a_7+\frac{1}{2}a_9+a_4
  \phantom{\frac{2m_{\eta^{(')}}^2}{(m_s+m_s)(m_b-m_d)}}\right.\right.\right.\nonumber\\
  &&\left.\left.-\frac{1}{2}a_{10}+\frac{m_{\eta^{(')}}^2}{m_s(m_b-m_d)}\big(a_6-\frac{1}{2}a_8\big)\Big(\frac{f^s_{\eta^{(')}}}{f^u_{\eta^{(')}}}-1\Big)r_{\eta^{(')}}\right)\right]
  X_{B^+\rightarrow \pi^+,\eta^{(')}_u}\nonumber\\
  &&-V_{tb}^*V_{td}\left(a_3-a_5+\frac{1}{2}a_7-\frac{1}{2}a_9\right)X_{B^+\rightarrow \pi^+,\eta^{(')}_s}+\left[V_{ub}^*V_{ud}a_1 \phantom{\frac{2m_{\pi^+}^2}{(m_u+m_d)(m_b-m_u)}}\right.\nonumber\\
  &&\left.\left.\hspace{-0.5ex}-V_{tb}^*V_{td}\left(a_4+a_{10}+\frac{2m_{\pi^+}^2}{(m_u+m_d)(m_b-m_u)}(a_6+a_8)\right)\right]X_{B^+\rightarrow \eta^{(')},\pi^+}\right\},
\end{eqnarray}
\begin{eqnarray}
  A(B^+\rightarrow \pi^+\omega)&=&\frac{G_F}{\sqrt{2}}\left\{\left[V_{ub}^*V_{ud}a_2-V_{tb}^*V_{td}\left(2a_3+2a_5+\frac{1}{2}a_7+\frac{1}{2}a_9+a_4-\frac{1}{2}a_{10}\right)\right]X_{B^+\rightarrow \pi^+,\omega}\right.\nonumber\\
  &&\left.+\left[V_{ub}^*V_{ud}a_1-V_{tb}^*V_{td}\left(a_4+a_{10} \phantom{\frac{2m_{\pi^+}^2}{(m_u+m_d)(m_b+m_u)}}\right.\right.\right.\nonumber\\
  &&\left.\left.\left.\hspace{-0.5ex}-\frac{2m_{\pi^+}^2}{(m_u+m_d)(m_b+m_u)}(a_6+a_8)\right)\right]X_{B^+\rightarrow \omega,\pi^+}\right\},
\end{eqnarray}
\begin{eqnarray}
  A(B^+\rightarrow \rho^+\eta^{(')})&=&\frac{G_F}{\sqrt{2}}\left\{\left[V_{ub}^*V_{ud}a_2-V_{tb}^*V_{td}\left(2a_3-2a_5-\frac{1}{2}a_7+\frac{1}{2}a_9+a_4\phantom{\frac{2m_{\eta^{(')}}^2}{(m_s+m_s)(m_b+m_d)}}\right.\right.\right.\nl
  &&\left.\left.-\frac{1}{2}a_{10}-\frac{m_{\eta^{(')}}^2}{m_s(m_b+m_d)}\big(a_6-\frac{1}{2}a_8\big)\Big(\frac{f^s_{\eta^{(')}}}{f^u_{\eta^{(')}}}-1\Big)r_{\eta^{(')}}\right)\right]X_{B^+\rightarrow \rho^+,\eta^{(')}_u}\nl
  &&-V_{tb}^*V_{td}\left(a_3-a_5+\frac{1}{2}a_7-\frac{1}{2}a_9\right)X_{B^+\rightarrow \rho^+,\eta^{(')}_s}\nl
  &&\left.\hspace{-0.3ex}+\Big[V_{ub}^*V_{ud}a_1-V_{tb}^*V_{td}\left(a_4+a_{10}\right)\Big]
  X_{B^+\rightarrow \eta^{(')},\rho^+}\phantom{\frac{2m_{\eta^{(')}}^2}{(m_s+m_s)(m_b+m_d)}}\hspace{-3.7cm}\right\},
\end{eqnarray}
\begin{eqnarray}
  A(B^+\rightarrow\pi^+ K^0)&=&-\frac{G_F}{\sqrt{2}}V_{tb}^*V_{ts}\left(a_4-\frac{1}{2}a_{10}+\frac{2m_{K^0}^2}{(m_s+m_d)(m_b-m_d)}(a_6-\frac{1}{2}a_8)\right)X_{B^+\rightarrow \pi^+,K^0},
\end{eqnarray}
\begin{eqnarray}
  A(B^+\rightarrow\rho^+ K^0)&=&-\frac{G_F}{\sqrt{2}}V_{tb}^*V_{ts}\left(a_4-\frac{1}{2}a_{10}-\frac{2m_{K^0}^2}{(m_s+m_d)(m_b+m_d)}(a_6-\frac{1}{2}a_8)\right)X_{B^+\rightarrow \rho^+,K^0},
\end{eqnarray}
\begin{eqnarray}
  A(B^+\rightarrow\pi^+\pi^0)&=&\frac{G_F}{\sqrt{2}}\left\{\left[V_{ub}^*V_{ud}a_2-V_{tb}^*V_{td}\left(\frac{3}{2}a_9-\frac{3}{2}a_7\right.\right.\right.\nl
  &&\left.\left.-a_4+\frac{1}{2}a_{10}-\frac{m_{\pi^0}^2}{m_d(m_b-m_d)}\big(a_6-\frac{1}{2}a_8\big)\right)\right]
  X_{B^+\rightarrow \pi^+,\pi^0}\nl
  &&\left.+\left[V_{ub}^*V_{ud}a_1-V_{tb}^*V_{td}\left(a_4+a_{10}\phantom{\frac{2m_{\pi^+}^2}{(m_u+m_d)(m_b-m_u)}}\right.\right.\right.\nl
  &&\left.\left.\left.\hspace{-0.5ex}+\frac{2m_{\pi^+}^2}{(m_u+m_d)(m_b-m_u)}(a_6+a_8)\right)\right]X_{B^+\rightarrow \pi^0,\pi^+}\right\},
\end{eqnarray}
\begin{eqnarray}
  A(B^+\rightarrow\pi^+\rho^0)&=&\frac{G_F}{\sqrt{2}}\left\{\left[V_{ub}^*V_{ud}a_2-V_{tb}^*V_{td}\left(-a_4+\frac{1}{2}a_{10}+\frac{3}{2}a_9+\frac{3}{2}a_7\right)\right]
  X_{B^+\rightarrow \pi^+,\rho^0}\right.\nl
  &&\left.+\left[V_{ub}^*V_{ud}a_1-V_{tb}^*V_{td}\left(a_4+a_{10}\phantom{\frac{2m_{\pi^+}^2}{(m_u+m_d)(m_b+m_u)}}\right.\right.\right.\nl
  &&\left.\left.\left.\hspace{-0.5ex}-\frac{2m_{\pi^+}^2}{(m_u+m_d)(m_b+m_u)}(a_6+a_8)\right)\right]X_{B^+\rightarrow \rho^0,\pi^+}\right\},
\end{eqnarray}
\begin{eqnarray}
  A(B^+\rightarrow\rho^+\pi^0)&=&\frac{G_F}{\sqrt{2}}\left\{\left[V_{ub}^*V_{ud}a_2-V_{tb}^*V_{td}\left(\frac{3}{2}a_9-\frac{3}{2}a_7-a_4+\frac{1}{2}a_{10}\right.\right.\right.\nl
  &&\left.\left.+\frac{m_{\pi^0}^2}{m_d(m_b+m_d)}\big(a_6-\frac{1}{2}a_8\big)\right)\right]
  X_{B^+\rightarrow \rho^+,\pi^0}\nl
  &&\left.+\left[V_{ub}^*V_{ud}a_1-V_{tb}^*V_{td}\left(a_4+a_{10}\right)\right]
  X_{B^+\rightarrow \pi^0,\rho^+}\phantom{\frac{1}{1}}\hspace{-1.5ex}\right\},
\end{eqnarray}
\begin{equation}
  A(B^+\rightarrow  \pi^+\phi)=-\frac{G_F}{\sqrt{2}}V_{tb}^*V_{td}\Big(a_3+a_5-\frac{1}{2}a_7-\frac{1}{2}a_9\Big)X_{B^+\rightarrow \pi^+,\phi},
\end{equation}
\begin{eqnarray}
  A(B^+\rightarrow\rho^+\rho^0)&=&\frac{G_F}{\sqrt{2}}\left\{\left[V_{ub}^*V_{ud}a_2-V_{tb}^*V_{td}\left(\frac{3}{2}a_9+\frac{3}{2}a_7-a_4+\frac{1}{2}a_{10}\right)\right]X_{B^+\rightarrow \rho^+,\rho^0}\right.\nl
  &&\left.+\Big[V_{ub}^*V_{ud}a_1-V_{tb}^*V_{td}\left(a_4+a_{10}\right)\Big]X_{B^+\rightarrow \rho^0,\rho^+}\phantom{\frac{1}{1}}\hspace{-1.5ex}\right\},
\end{eqnarray}
\begin{eqnarray}
  A(B^+\rightarrow\rho^+\omega)&=&\frac{G_F}{\sqrt{2}}\left\{\left[V_{ub}^*V_{ud}a_2-V_{tb}^*V_{td}\left(2a_3+2a_5+\frac{1}{2}a_7+\frac{1}{2}a_9+a_4-\frac{1}{2}a_{10}\right)\right]X_{B^+\rightarrow \rho^+,\omega}\right.\nl
  &&\left.+\Big[V_{ub}^*V_{ud}a_1-V_{tb}^*V_{td}\left(a_4+a_{10}\right)\Big]X_{B^+\rightarrow \omega,\rho^+}\phantom{\frac{1}{1}}\hspace{-1.5ex}\right\},
\end{eqnarray}
\begin{equation}
A(B^0\rightarrow D^- \pi^+)=\frac{G_F}{\sqrt{2}}V_{cb}^*V_{ud}a_1X_{B^0\rightarrow D^-,\pi^+},
\end{equation}
\begin{equation}
A(B^0\rightarrow D^- K^+)=\frac{G_F}{\sqrt{2}}V_{cb}^*V_{us}a_1X_{B^0\rightarrow D^-,K^+},
\end{equation}
\begin{eqnarray}
A(B^0\rightarrow \pi^- K^+)&=&\frac{G_F}{\sqrt{2}}\left[V_{ub}^*V_{us}a_1-V_{tb}^*V_{ts}\left(a_4+a_{10}
\phantom{\frac{2m_{K^+}^2}{(m_u+m_s)(m_b-m_u)}}\right.\right.\nl
&&\left.\left.+\frac{2m_{K^+}^2}{(m_u+m_s)(m_b-m_u)}(a_6+a_8)\right)\right]X_{B^0\rightarrow \pi^-,K^+},
\end{eqnarray}
\begin{eqnarray}
A(B^0\rightarrow \pi^- \pi^+)&=&\frac{G_F}{\sqrt{2}}\left[V_{ub}^*V_{ud}a_1-V_{tb}^*V_{td}\left(a_4+a_{10}
\phantom{\frac{2m_{\pi^+}^2}{(m_u+m_d)(m_b-m_u)}}\right.\right.\nl
&&\left.\left.+\frac{2m_{\pi^+}^2}{(m_u+m_d)(m_b-m_u)}(a_6+a_8)\right)\right]X_{B^0\rightarrow \pi^-,\pi^+},
\end{eqnarray}
\begin{eqnarray}
A(B^0\rightarrow \pi^0 \pi^0)&=&2\frac{G_F}{\sqrt{2}}\left[V_{ub}^*V_{ud}a_2-V_{tb}^*V_{td}\left(\frac{3}{2}a_9-\frac{3}{2}a_7-a_4+\frac{1}{2}a_{10}\right.\right.\nl
&&\left.\left.-\frac{m_{\pi^0}^2}{(m_d(m_b-m_d)}\big(a_6-\frac{1}{2}a_8\big)\right)\right]X_{B^0\rightarrow \pi^0,\pi^0},
\end{eqnarray}
\begin{eqnarray}
  A(B^0\rightarrow \pi^0\eta^{(')})&=&\frac{G_F}{\sqrt{2}}\left\{\left[V_{ub}^*V_{ud}a_2-V_{tb}^*V_{td}\left(2a_3-2a_5-\frac{1}{2}a_7+\frac{1}{2}a_9+a_4-\frac{1}{2}a_{10}\right.\right.\right.\nl
  &&\left.\left.+\frac{m_{\eta^{(')}}^2}{m_s(m_b-m_d)}\big(a_6-\frac{1}{2}a_8\big)\Big(\frac{f^s_{\eta^{(')}}}{f^u_{\eta^{(')}}}-1\Big)r_{\eta^{(')}}\right)\right]
  X_{B^0\rightarrow \pi^0,\eta^{(')}_u}\nl
  &&-V_{tb}^*V_{td}\left(a_3-a_5+\frac{1}{2}a_7-\frac{1}{2}a_9\right)X_{B^0\rightarrow \pi^0,\eta^{(')}_s}\nl
  &&+\left[V_{ub}^*V_{ud}a_2-V_{tb}^*V_{td}\left(\frac{3}{2}a_9-\frac{3}{2}a_7-a_4+\frac{1}{2}a_{10}\right.\right.\nl
  &&\left.\left.\left.-\frac{m_{\pi^{0}}^2}{m_d(m_b-m_d)}\big(a_6-\frac{1}{2}a_8\big)\right)\right]
  X_{B^0\rightarrow \eta^{(')},\pi^0}\right\},
\end{eqnarray}
\begin{eqnarray}
  A(B^0\rightarrow\eta\eta)&=&2\frac{G_F}{\sqrt{2}}\left\{\left[\phantom{\frac{2m_{\eta}^2}{(m_s+m_s)(m_b-m_d)}}\hspace{-3.5cm}V_{ub}^*V_{ud}a_2-V_{tb}^*V_{td}\left(2a_3-2a_5-\frac{1}{2}a_7+\frac{1}{2}a_9+a_4-\frac{1}{2}a_{10} \right.\right.\right.\nonumber\\ &&\left.\left.+\frac{2m_{\eta}^2}{(m_s+m_s)(m_b-m_d)}\big(a_6-\frac{1}{2}a_8\big)\Big(\frac{f^s_{\eta}}{f^u_{\eta}}-1\Big)r_{\eta}\right)\right]X_{B^0\rightarrow \eta,\eta_u}\nonumber\\
   &&\left.-V_{tb}^*V_{td}\left(a_3-a_5+\frac{1}{2}a_7-\frac{1}{2}a_9\right)X_{B^0\rightarrow \eta,\eta_s}\phantom{\frac{m_{\eta}^2}{m_s(m_b-m_d)}}\hspace{-2.4cm}\right\},
\end{eqnarray}
\begin{eqnarray}
  A(B^0\rightarrow\eta'\eta')&=&2\frac{G_F}{\sqrt{2}}\left\{\left[\phantom{\frac{2m_{\eta'}^2}{(m_s+m_s)(m_b-m_d)}}\hspace{-3.5cm}V_{ub}^*V_{ud}a_2-V_{tb}^*V_{td}\left(2a_3-2a_5-\frac{1}{2}a_7+\frac{1}{2}a_9+a_4-\frac{1}{2}a_{10} \right.\right.\right.\nonumber\\ &&\left.\left.+\frac{m_{\eta'}^2}{m_s(m_b-m_d)}\big(a_6-\frac{1}{2}a_8\big)\Big(\frac{f^s_{\eta'}}{f^u_{\eta'}}-1\Big)r_{\eta'}\right)\right]X_{B^0\rightarrow \eta',\eta'_u}\nonumber\\
   &&\left.-V_{tb}^*V_{td}\left(a_3-a_5+\frac{1}{2}a_7-\frac{1}{2}a_9\right)X_{B^0\rightarrow \eta',\eta'_s}\phantom{\frac{2m_{\eta'}^2}{(m_s+m_s)(m_b-m_d)}}\hspace{-3.5cm}\right\},
\end{eqnarray}
\begin{equation}
A(B^0\rightarrow D^{*-} \pi^+)=\frac{G_F}{\sqrt{2}}V_{cb}^*V_{ud}a_1X_{B^0\rightarrow D^{*-},\pi^+},
\end{equation}
\begin{equation}
A(B^0\rightarrow D^- \rho^+)=\frac{G_F}{\sqrt{2}}V_{cb}^*V_{ud}a_1X_{B^0\rightarrow D^-,\rho^+},
\end{equation}
\begin{eqnarray}
  A(B^0\rightarrow \rho^0\pi^0)&=&\frac{G_F}{\sqrt{2}}\left\{\left[V_{ub}^*V_{ud}a_2-V_{tb}^*V_{td}\left(\frac{3}{2}a_7+\frac{3}{2}a_9-a_4+\frac{1}{2}a_{10}\right)\right]X_{B^0\rightarrow \pi^0,\rho^0}\right.\nonumber\\
 &&+\left[V_{ub}^*V_{ud}a_2-V_{tb}^*V_{td}\left(\frac{3}{2}a_9-\frac{3}{2}a_7-a_4+\frac{1}{2}a_{10}\right.\right.\nonumber\\
  &&\left.\left.\left.\hspace{-0.5ex}+\frac{m_{\pi^0}^2}{m_d(m_b+m_d)}\big(a_6-\frac{1}{2}a_8\big)\right)\right]X_{B^0\rightarrow\rho^0,\pi^0}\right\},
\end{eqnarray}
\begin{eqnarray}
  A(B^0\rightarrow \omega\pi^0)&=&\frac{G_F}{\sqrt{2}}\left\{\left[V_{ub}^*V_{ud}a_2-V_{tb}^*V_{td}\left(2a_3+2a_5+\frac{1}{2}a_7+\frac{1}{2}a_9+a_4-\frac{1}{2}a_{10}\right)\right]X_{B^0\rightarrow \pi^0,\omega}\right.\nl
 &&+\left[V_{ub}^*V_{ud}a_2-V_{tb}^*V_{td}\left(\frac{3}{2}a_9-\frac{3}{2}a_7-a_4+\frac{1}{2}a_{10}\right.\right.\nl
 &&\left.\left.\left.\hspace{-0.5ex}+\frac{m_{\pi^0}^2}{m_d(m_b+m_d)}\big(a_6-\frac{1}{2}a_8\big)\right)\right]X_{B^0\rightarrow\omega,\pi^0}\right\},
\end{eqnarray}
\begin{eqnarray}
A(B^0\rightarrow \rho^- K^+)&=&\frac{G_F}{\sqrt{2}}\left[V_{ub}^*V_{us}a_1-V_{tb}^*V_{ts}\left(a_4+a_{10}\phantom{\frac{2m_{K^+}^2}{(m_s+m_u)(m_b+m_u)}}\right.\right.\nl
&&\left.\left.-\frac{2m_{K^+}^2}{(m_s+m_u)(m_b+m_u)}(a_6+a_8)\right)\right]X_{B^0\rightarrow \rho^-,K^+},
\end{eqnarray}
\begin{equation}
A(B^0\rightarrow \pi^- K^{*+})=\frac{G_F}{\sqrt{2}}\Big[V_{ub}^*V_{us}a_1-V_{tb}^*V_{ts}\left(a_4+a_{10}\right)\Big]X_{B^0\rightarrow \pi^-,K^{*+}},
\end{equation}
\begin{eqnarray}
A(B^0\rightarrow \rho^- \pi^+)&=&\frac{G_F}{\sqrt{2}}\left[V_{ub}^*V_{ud}a_1-V_{tb}^*V_{td}\left(a_4+a_{10}\phantom{\frac{2m_{\pi^+}^2}{(m_u+m_d)(m_b+m_u)}}\right.\right.\nl
&&\left.\left.-\frac{2m_{\pi^+}^2}{(m_u+m_d)(m_b+m_u)}(a_6+a_8)\right)\right]X_{B^0\rightarrow \rho^-,\pi^+},
\end{eqnarray}
\begin{equation}
A(B^0\rightarrow D^- K^{*+})=\frac{G_F}{\sqrt{2}}V_{cb}^*V_{us}a_1X_{B^0\rightarrow D^-,K^{*+}},
\end{equation}
\begin{equation}
A(B^0\rightarrow D^{*-} K^{+})=\frac{G_F}{\sqrt{2}}V_{cb}^*V_{us}a_1X_{B^0\rightarrow  D^{*-},K^{+}},
\end{equation}
\begin{eqnarray}
  A(B^0\rightarrow \eta'\eta)&=&\frac{G_F}{\sqrt{2}}\left\{\phantom{\frac{2m_{\eta}^2}{(m_s+m_s)(m_b-m_d)}}\hspace{-3.5cm}\left[V_{ub}^*V_{ud}a_2-V_{tb}^*V_{td}\left(2a_3-2a_5-\frac{1}{2}a_7+\frac{1}{2}a_9+a_4-\frac{1}{2}a_{10}\right.\right.\right.\nl
   &&\left.\left.+\frac{m_{\eta}^2}{m_s(m_b-m_d)}\big(a_6-\frac{1}{2}a_8\big)\Big(\frac{f^s_{\eta}}{f^u_{\eta}}-1\Big)r_{\eta}\right)\right]
   X_{B^0\rightarrow \eta^{'},\eta_u}\nl
  &&-V_{tb}^*V_{td}\left(a_3-a_5+\frac{1}{2}a_7-\frac{1}{2}a_9\right)X_{B^0\rightarrow  \eta',\eta_s}\nl
  &&+\left[V_{ub}^*V_{ud}a_2-V_{tb}^*V_{td}\left(2a_3-2a_5-\frac{1}{2}a_7+\frac{1}{2}a_9+a_4-\frac{1}{2}a_{10}\right.\right.\nl
  &&\left.\left.\hspace{-0.5ex}+\frac{m_{\eta^{'}}^2}{m_s(m_b-m_d)}\big(a_6-\frac{1}{2}a_8\big)\Big(\frac{f^s_{\eta^{'}}}{f^u_{\eta^{'}}}-1\Big)r_{\eta^{'}}\right)\right]
  X_{B^0\rightarrow \eta,\eta^{'}_u}\nl
  &&\left.\hspace{-0.5ex}-V_{tb}^*V_{td}\left(a_3-a_5+\frac{1}{2}a_7-\frac{1}{2}a_9\right)X_{B^0\rightarrow  \eta,\eta'_s}\phantom{\frac{2m_{\eta}^2}{(m_s+m_s)(m_b-m_d)}}\hspace{-3.5cm}\right\},
\end{eqnarray}
\begin{eqnarray}
  A(B^0\rightarrow\rho^0\eta^{(')})&=&\frac{G_F}{\sqrt{2}}\left\{\left[V_{ub}^*V_{ud}a_2-V_{tb}^*V_{td}\left(2a_3-2a_5-\frac{1}{2}a_7+\frac{1}{2}a_9+a_4-\frac{1}{2}a_{10}\right.\right.\right.\nl
   &&\left.\left.-\frac{m_{\eta^{(')}}^2}{m_s(m_b+m_d)}\big(a_6-\frac{1}{2}a_8\big)\Big(\frac{f^s_{\eta^{(')}}}{f^u_{\eta^{(')}}}-1\Big)r_{\eta^{(')}}\right)\right]X_{B^0\rightarrow\rho^0,\eta^{(')}_u}\nl
  &&-V_{tb}^*V_{td}\left(a_3-a_5+\frac{1}{2}a_7-\frac{1}{2}a_9\right)X_{B^0\rightarrow \rho^0,\eta^{(')}_s}\nl
  &&\left.+\left[V_{ub}^*V_{ud}a_2-V_{tb}^*V_{td}\left(\frac{3}{2}a_7+\frac{3}{2}a_9-a_4+\frac{1}{2}a_{10}\right)\right]
  X_{B^0\rightarrow \eta^{(')},\rho^0}\right\},
\end{eqnarray}
\begin{eqnarray}
  A(B^0\rightarrow\omega\eta^{(')})&=&\frac{G_F}{\sqrt{2}}\left\{\left[\phantom{\frac{2m_{\eta^{(')}}^2}{(m_s+m_s)(m_b+m_d)}}\hspace{-3.5cm}V_{ub}^*V_{ud}a_2-V_{tb}^*V_{td}\left(2a_3-2a_5-\frac{1}{2}a_7+\frac{1}{2}a_9+a_4-\frac{1}{2}a_{10}\right.\right.\right.\nl
  &&\left.\left.-\frac{m_{\eta^{(')}}^2}{m_d(m_b+m_d)}\big(a_6-\frac{1}{2}a_8\big)\Big(\frac{f^s_{\eta^{(')}}}{f^u_{\eta^{(')}}}-1\Big)r_{\eta^{(')}}\right)\right]X_{B^0\rightarrow\omega,\eta^{(')}_u}\nl
  &&-V_{tb}^*V_{td}\left(a_3-a_5+\frac{1}{2}a_7-\frac{1}{2}a_9\right)X_{B^0\rightarrow \omega,\eta^{(')}_s}+\left[V_{ub}^*V_{ud}a_2\phantom{\frac{1}{1}}\right.\nl
  &&\left.\left.-V_{tb}^*V_{td}\left(2a_3+2a_5+\frac{1}{2}a_7+\frac{1}{2}a_9+a_4-\frac{1}{2}a_{10}\right)\right]
  X_{B^0\rightarrow \eta^{(')},\omega}\phantom{\frac{2m_{\eta^{(')}}^2}{(m_s+m_s)(m_b+m_d)}}\hspace{-3.5cm}\right\},
\end{eqnarray}
\begin{equation}
A(B^0\rightarrow \pi^-\rho^{+})=\frac{G_F}{\sqrt{2}}\Big[V_{ub}^*V_{ud}a_1-V_{tb}^*V_{td}\left(a_4+a_{10}\right)\Big]X_{B^0\rightarrow \pi^-,\rho^{+}},
\end{equation}
\begin{equation}
A(B^0\rightarrow \rho^- \rho^+)=\frac{G_F}{\sqrt{2}}\Big[V_{ub}^*V_{ud}a_1-V_{tb}^*V_{td}\left(a_4+a_{10}\right)\Big]X_{B^0\rightarrow \rho^-,\rho^+},
\end{equation}
\begin{equation}
A(B^0\rightarrow \rho^0 \rho^0)=2\frac{G_F}{\sqrt{2}}\left[V_{ub}^*V_{ud}a_2-V_{tb}^*V_{td}\left(\frac{3}{2}a_9+\frac{3}{2}a_7-a_4+\frac{1}{2}a_{10}\right)\right]X_{B^0\rightarrow \rho^0,\rho^0},
\end{equation}

\begin{equation}
A(B^0\rightarrow\omega \omega)=2\frac{G_F}{\sqrt{2}}\left[V_{ub}^*V_{ud}a_2-V_{tb}^*V_{td}\left(2a_3+2a_5+\frac{1}{2}a_7+\frac{1}{2}a_9+a_4-\frac{1}{2}a_{10}\right)\right]X_{B^0\rightarrow \omega,\omega},
\end{equation}
\begin{eqnarray}
A(B^0\rightarrow\omega\rho^0)&=&\frac{G_F}{\sqrt{2}}\left\{\left[V_{ub}^*V_{ud}a_2-V_{tb}^*V_{td}\left(2a_3+2a_5+\frac{1}{2}a_7+\frac{1}{2}a_9+a_4-\frac{1}{2}a_{10}\right)\right]X_{B^0\rightarrow \rho^0,\omega}\right.\nl
&&\left.+\left[V_{ub}^*V_{ud}a_2-V_{tb}^*V_{td}\left(\frac{3}{2}a_7+\frac{3}{2}a_9-a_4+\frac{1}{2}a_{10}\right)\right]X_{B^0\rightarrow \omega,\rho^0}\right\},
\end{eqnarray}
\begin{equation}
A(B^0\rightarrow D^{*-} \rho^+)=\frac{G_F}{\sqrt{2}}V_{cb}^*V_{ud}a_1X_{B^0\rightarrow D^{*-},\rho^+},
\end{equation}
\begin{equation}
A(B^0\rightarrow D^{*-} K^{*+})=\frac{G_F}{\sqrt{2}}V_{cb}^*V_{us}a_1X_{B^0\rightarrow D^{*-},K^{*+}},
\end{equation}
\begin{equation}
A(B_s\rightarrow D_s^- \pi^+)=\frac{G_F}{\sqrt{2}}V_{cb}^*V_{ud}a_1X_{B_s\rightarrow D_s^-,\pi^+},
\end{equation}
\begin{equation}
A(B_s\rightarrow D_s^- K^+)=\frac{G_F}{\sqrt{2}}V_{cb}^*V_{us}a_1X_{B_s\rightarrow D_s^-,K^+},
\end{equation}
\begin{eqnarray}
A(B_s\rightarrow \pi^+ K^-)&=&\frac{G_F}{\sqrt{2}}\left[V_{ub}^*V_{ud}a_1-V_{tb}^*V_{td}\left(a_4+a_{10}\phantom{\frac{2m_{\pi^+}^2}{(m_u+m_d)(m_b-m_u)}}\right.\right.\nl
&&\left.\left.+\frac{2m_{\pi^+}^2}{(m_u+m_d)(m_b-m_u)}(a_6+a_8)\right)\right]X_{B_s\rightarrow K^-,\pi^+},
\end{eqnarray}
\begin{eqnarray}
A(B_s\rightarrow K^+ K^-)&=&\frac{G_F}{\sqrt{2}}\left[V_{ub}^*V_{us}a_1-V_{tb}^*V_{ts}\left(a_4+a_{10}\phantom{\frac{2m_{K^+}^2}{(m_u+m_s)(m_b-m_u)}}\right.\right.\nl
&&\left.\left.+\frac{2m_{K^+}^2}{(m_u+m_s)(m_b-m_u)}(a_6+a_8)\right)\right]X_{B_s\rightarrow K^-,K^+},
\end{eqnarray}
\begin{equation}
A(B_s\rightarrow D_s^- \rho^+)=\frac{G_F}{\sqrt{2}}V_{cb}^*V_{ud}a_1X_{B_s\rightarrow D_s^-,\rho^+},
\end{equation}
\begin{equation}
A(B_s\rightarrow D_s^{*-} \pi^+)=\frac{G_F}{\sqrt{2}}V_{cb}^*V_{ud}a_1X_{B_s\rightarrow D_s^{*-},\pi^+},
\end{equation}
\begin{equation}
A(B_s\rightarrow D_s^{*-} K^+)=\frac{G_F}{\sqrt{2}}V_{cb}^*V_{us}a_1X_{B_s\rightarrow D_s^{*-},K^+},
\end{equation}
\begin{equation}
A(B_s\rightarrow D_s^{-} K^{*+})=\frac{G_F}{\sqrt{2}}V_{cb}^*V_{us}a_1X_{B_s\rightarrow D_s^{-},K^{*+}},
\end{equation}
\begin{eqnarray}
A(B_s\rightarrow K^{*-} \pi^+)&=&\frac{G_F}{\sqrt{2}}\left[V_{ub}^*V_{ud}a_1-V_{tb}^*V_{td}\left(a_4+a_{10}\phantom{\frac{2m_{\pi^+}^2}{(m_u+m_d)(m_b+m_u)}}\right.\right.\nl
&&\left.\left.-\frac{2m_{\pi^+}^2}{(m_u+m_d)(m_b+m_u)}(a_6+a_8)\right)\right]X_{B_s\rightarrow K^{*-},\pi^+},
\end{eqnarray}
\begin{eqnarray}
A(B_s\rightarrow K^{*-} K^+)&=&\frac{G_F}{\sqrt{2}}\left[V_{ub}^*V_{us}a_1-V_{tb}^*V_{ts}\left(a_4+a_{10}\phantom{\frac{2m_{K^+}^2}{(m_u+m_s)(m_b+m_u)}}\right.\right.\nl
&&\left.\left.-\frac{2m_{K^+}^2}{(m_u+m_s)(m_b+m_u)}(a_6+a_8)\right)\right]
X_{B_s\rightarrow K^{*-},K^+},
\end{eqnarray}
\begin{equation}
A(B_s\rightarrow  K^{-}K^{*+})=\frac{G_F}{\sqrt{2}}\Big[V_{ub}^*V_{us}a_1-V_{tb}^*V_{ts}(a_4+a_{10})\Big]X_{B_s\rightarrow  K^{-},K^{*+}},
\end{equation}
\begin{equation}
A(B_s\rightarrow D_s^{*-} K^{*+})=\frac{G_F}{\sqrt{2}}V_{cb}^*V_{us}a_1X_{B_s\rightarrow D_s^{*-},K^{*+}},
\end{equation}
\begin{equation}
A(B_s\rightarrow D_s^{*-}\rho^+)=\frac{G_F}{\sqrt{2}}V_{cb}^*V_{ud}a_1X_{B_s\rightarrow D_s^{*-},\rho^+}.
\end{equation}


\end{document}